\documentclass[10pt, conference, compsocconf]{IEEEtran}

\usepackage{cite}
\ifCLASSINFOpdf
  \usepackage[pdftex]{graphicx}
\else
  \usepackage[dvips]{graphicx}
\fi
\usepackage[cmex10]{amsmath}
\usepackage{mdwmath}
\usepackage{mdwtab}
\usepackage[tight,footnotesize]{subfigure}
\usepackage{url}

\usepackage{balance}
\def\IEEEbibitemsep{3pt plus .5pt}

\usepackage{amsfonts}
\usepackage{comment}
\usepackage{tabularx}
\usepackage{algorithm, algpseudocode}
\usepackage{xspace}
\usepackage{multirow}
\usepackage{threeparttable}
\usepackage{mathtools}
\usepackage{booktabs}
\usepackage{xfrac}

\newcommand{\NAME}{ARK\xspace}

\newcommand{\dnum}{$\mathtt{dnum}$\xspace}
\newcommand{\dnums}{$\mathtt{dnum}$s\xspace}
\newcommand{\evk}{$\mathbf{evk}$\xspace}
\newcommand{\evks}{$\mathbf{evk}$s\xspace}
\newcommand{\amort}{$\textbf{T}_{A.S.}$\xspace}

\begin{document}

\title{ARK: Fully Homomorphic Encryption Accelerator with Runtime Data Generation and Inter-Operation Key Reuse}

\author{\IEEEauthorblockN{Jongmin Kim\IEEEauthorrefmark{1}\IEEEauthorrefmark{2},
Gwangho Lee\IEEEauthorrefmark{1}\IEEEauthorrefmark{2},
Sangpyo Kim\IEEEauthorrefmark{2}, 
Gina Sohn\IEEEauthorrefmark{2},
Minsoo Rhu\IEEEauthorrefmark{3},
John Kim\IEEEauthorrefmark{3},
Jung Ho Ahn\IEEEauthorrefmark{2}}
\IEEEauthorblockA{\IEEEauthorrefmark{2}Seoul National University, Seoul, South Korea\\
\{jongmin.kim, g\_corey, vnb987, gina.gemini, gajh\}@snu.ac.kr}
\IEEEauthorblockA{\IEEEauthorrefmark{3}KAIST, Daejeon, South Korea\\
\{mrhu, jjk12\}@kaist.edu}}

\maketitle

\def\thefootnote{}\footnotetext{\IEEEauthorrefmark{1}Jongmin Kim and Gwangho Lee are co-first authors.}
\def\thefootnote{\arabic{footnote}}

\begin{abstract}
Homomorphic Encryption (HE) is one of the most promising post-quantum cryptographic schemes that enable privacy-preserving computation on servers.
However, noise accumulates as we perform operations on HE-encrypted data, restricting the number of possible operations.
Fully HE (FHE) removes this restriction by introducing the bootstrapping operation, which refreshes the data; however, FHE schemes are highly memory-bound.
Bootstrapping, in particular, requires loading GBs of evaluation keys and plaintexts from off-chip memory, which makes FHE acceleration fundamentally bottlenecked by the off-chip memory bandwidth.

In this paper, we propose \NAME, an \textbf{A}ccelerator for FHE with \textbf{R}untime data generation and inter-operation \textbf{K}ey reuse. \NAME enables practical FHE workloads with a novel algorithm-architecture co-design to accelerate bootstrapping.
We first eliminate the off-chip memory bandwidth bottleneck through runtime data generation and inter-operation key reuse.
This approach enables \NAME to fully exploit on-chip memory by substantially reducing the size of the working set.
On top of such algorithmic enhancements, we build \NAME microarchitecture that minimizes on-chip data movement through an efficient, alternating data distribution policy based on the data access patterns and a streamlined dataflow organization of the tailored functional units\,---\,including base conversion, number-theoretic transform, and automorphism units.
Overall, our co-design effectively handles the heavy computation and data movement overheads of FHE, drastically reducing the cost of HE operations, including bootstrapping.
\end{abstract}


\IEEEpeerreviewmaketitle

\section{Introduction}
\label{sec:1_introduction}

Homomorphic encryption (HE) is a cryptographic method that enables third parties to process data blindly.
HE is regarded as a game-changer in privacy-preserving cloud computing because HE does not leak information about the data to anyone other than the owner, allowing secure offloading of computation to untrusted servers.

HE schemes are based on the learning with errors~\cite{jacm-2009-lwe} paradigm, a computational problem using noise for security.
Noise accumulates as we continuously apply computation to encrypted data, which can eventually explode and make the data impossible to decrypt.
Leveled HE (LHE) schemes only allow a restricted number of operations (ops) to prevent noise explosion, thus only supporting simple workloads~\cite{siam-2014-bfv1, iacr-2015-guide}.
An HE op called \emph{bootstrapping} can be used to lower the noise level and allow an unbounded number of ops over encrypted data.
Due to its significance, this class of HE schemes capable of bootstrapping is referred to as \emph{fully HE} (FHE)~\cite{stoc-2009-gentry-fhe}.
Among numerous FHE schemes~\cite{iacr-2012-bfv2, siam-2014-bfv1, toct-2014-bgv, jc-2020-tfhe, eurocrypt-2015-fhew}, CKKS~\cite{asia-2017-ckks}, our primary target FHE scheme, is gaining popularity because it supports real arithmetic and is thus more suitable for machine learning (ML) services.

Despite its potential, however, the high memory and computational costs of HE are preventing its wide adoption.
When data is encrypted using an HE scheme, the size expands by more than an order of magnitude, and an op as simple as multiplication transforms into a complex sequence of ops.
Prior works have tried to mitigate such overhead by using CPU extensions~\cite{access-2021-demystify, wahc-2021-hexl}, GPUs~\cite{tches-2021-100x, access-2020-privft, tetc-2019-bfv, tches-2018-fv}, FPGAs~\cite{asplos-2020-heax, hpca-2019-roy, fccm-2020-sunwoong-ntt, reconfig-2019-sunwoong-modmult}, and ASICs~\cite{micro-2021-f1, hpca-2021-cheetah}.
Recent proposals suggesting accelerated ASIC/FPGA designs, F1~\cite{micro-2021-f1} in particular, have proven successful for LHE schemes.

However, the parameters used in these designs were not large enough to effectively support CKKS bootstrapping.
Bootstrapping is an expensive op, dominating the execution time of complex FHE workloads~\cite{access-2022-resnet20, tches-2021-100x, icml-2022-resnet}.
Moreover, FHE schemes supporting bootstrapping must use large parameters~\cite{korean-2022-cheon-practical, eurocrypt-2021-efficient}, which correspondingly increase the data size.
In fact, the working set of bootstrapping can reach several GBs, which must be fetched from off-chip memory due to limited on-chip memory capacity~\cite{cse-2022-h100, vlsit-2018-samsung}.
Given such constraints of FHE schemes, previous hardware accelerators are severely bottlenecked by the low off-chip memory bandwidth.

In this paper, we present a set of algorithmic enhancements that address the off-chip memory bandwidth bottleneck of FHE workloads: \emph{minimum key-switching} and \emph{on-the-fly limb extension}.
The key benefits of our proposed algorithms are twofold.
First, our proposed algorithms drastically reduce the memory bottleneck of FHE, eliminating 88\% of off-chip memory access in evaluating a major op composing bootstrapping. Second, our proposal enables the deployment of massive computational logic and on-chip memory, large enough to hold the reduced working set on-chip, a feature lacking in conventional systems.

We design \NAME, an FHE accelerator that effectively exploits the increased arithmetic intensity enabled by our proposed algorithmic enhancements, addressing new computational and data movement requirements arising from FHE support.
\NAME features novel functional units (FUs) tailored to the primary functions of FHE, including base conversion (BConv), number-theoretic transform (NTT), and automorphism.
The FUs lower the bandwidth pressure on register files (RFs) and minimize data movement.
For seamless operation across the FUs, \NAME utilizes a data distribution policy that adapts to two data access patterns of the primary functions and simplifies the dataflow by splitting the chip into two logical regions.

By co-designing the algorithms and the architecture, \NAME outperforms the state-of-the-art HE accelerator~\cite{micro-2021-f1} in multiplicative throughput~\cite{tches-2021-100x} by 2,353$\times$ and in logistic regression training~\cite{aaai-2019-helr} by 18$\times$.
\NAME also enables a real-time CNN inference on encrypted data; \NAME performs an inference with the ResNet-20 model in 0.125 seconds, 18,214$\times$ faster than the baseline CPU implementation~\cite{icml-2022-resnet}.
\NAME is sized 418.3mm\textsuperscript{2} and consumes up to 281.3W of power.

In summary, our major contributions are as follows:
\begin{itemize}
  \item We analyze CKKS bootstrapping in detail and identify the memory bottleneck in hardware acceleration of FHE induced by excessive single-use data during bootstrapping.
  \item We devise two algorithms to mitigate the memory bottleneck of FHE by substantially reducing the single-use data in bootstrapping. These algorithms break new ground specifically for domain-specific architectures by overcoming the performance bound imposed by the memory bottleneck.
  \item We design \NAME, an algorithm-architecture co-designed accelerator for FHE. We deploy specialized FUs that minimize on-chip data movement on \NAME. \NAME adapts data and job distribution based on the data access patterns and fixes dataflow for streamlined execution across the chip.
\end{itemize}

\section{Background}
\label{sec:2_background}

We explain the pertinent details of HE with an emphasis on CKKS.
We adopt notations and terminologies from \cite{arxiv-2021-does, rsa-2020-better}.
Parameters and notations are summarized in Table \ref{table:notation}.

\subsection{Homomorphic encryption (HE)}
\label{sec:2_1_he}

HE is an encryption method that allows computation on encrypted data.
It is based on a computational problem called ring learning with errors (RLWE), which is known to be difficult to solve even for quantum computers~\cite{2021-standard}.
There are a variety of HE schemes that support different data types, such as integers~\cite{toct-2014-bgv, siam-2014-bfv1, iacr-2012-bfv2}, boolean~\cite{jc-2020-tfhe, eurocrypt-2015-fhew}, and fixed-point complex (real) numbers~\cite{asia-2017-ckks}.
In particular, CKKS~\cite{asia-2017-ckks}, our primary target HE scheme, supports the encryption of fixed-point complex  vectors, making it suitable for numerous real-world applications such as machine learning.

In CKKS, a \emph{message} is an unencrypted vector of length $n$.
Each entry of a message vector is called a \emph{slot}, which can contain a complex number.
\emph{Encryption} of a message consists of two steps.
First, we transform a message $\textbf{m}$ into a polynomial $\mathtt{P}_{\textbf{m}}$ in a cyclotomic polynomial ring $\mathrm{R}_Q\! =\! \mathbb{Z}_Q[X] / (X^N\! + 1)$, which is a degree-$(N\! - 1)$ polynomial with coefficients in $ \mathbb{Z}_Q\! =\! \{0, 1, \cdots, Q\! -\! 1\} $, the set of integers modulo $Q$; we treat a polynomial as a vector in $ \mathbb{Z}_Q^N$ where the entries are the polynomial's coefficients.
The maximum number of slots included in a single message $\textbf{m}$ is determined by $N$ as $n \leq \sfrac{N}{2}$.
We perform this transformation by computing the Inverse Discrete Fourier Transform (IDFT) of $\textbf{m}$, multiplying a scale factor $\Delta$, and rounding the result:
\begin{equation} \label{eq:encode}
        \mathtt{P}_\mathbf{m} \simeq \Delta \cdot \text{IDFT}(\mathbf{m})
\end{equation}
Then, $\mathtt{P}_{\textbf{m}}$ is converted into a \emph{ciphertext} $[\![ \mathbf{m} ]\!]$ based on Eq.~\ref{eq:encryption}.
$\mathtt{A}_{\textbf{m}}$ is a randomly sampled polynomial, $\mathtt{S}$ is a secret key polynomial, and $ \mathtt{E} $ is a small random error polynomial.
\begin{equation} \label{eq:encryption}
    [\![ \mathbf{m} ]\!]= ( \mathtt{B}_{\textbf{m}}, \mathtt{A}_{\textbf{m}} ),\ \ \ \ \mathtt{B}_\mathbf{m} =  \mathtt{A}_\mathbf{m} * \mathtt{S} + \mathtt{P}_\mathbf{m} + \mathtt{E}
\end{equation}

\emph{Decryption} is the inverse of the encryption process.
We first recover the polynomial $\mathtt{P}_\mathbf{m}$ from $[\![ \mathbf{m} ]\!]$ using the secret key $\mathtt{S}$. Then, we apply DFT as shown in the following equation.
\begin{equation} \label{eq:decode}
    \mathbf{m} \simeq  {\Delta}^{-1} \cdot \text{DFT}(\mathtt{P}_\mathbf{m}+\mathtt{E})={\Delta}^{-1} \cdot \text{DFT}( \mathtt{B}_{\textbf{m}}-\mathtt{A}_{\textbf{m}}*\mathtt{S} )
\end{equation}
Due to the random error polynomial $\mathtt{E}$ included in the encryption process, the decrypted result is not identical to $\textbf{m}$.
However, we can consider the decrypted result as $\textbf{m}$ because $\mathtt{E}$ is a polynomial with small coefficients.

\setlength{\tabcolsep}{4pt}
\renewcommand{\arraystretch}{1.2}
\begin{table}[t]
  \caption{CKKS parameters and notations}
  \label{table:notation}
  \centering
  \begin{tabularx}{\columnwidth}{lX}
    \toprule
    \textbf{Params}                     & \textbf{Description}\\
    \midrule
    $N$                                 & Degree. Length of a plaintext polynomial.\\
    $n$                                 & Length of a message. $ n \leq N/2 $.\\
    $Q$                                 & Polynomial modulus.\\
    $P$                                 & Special modulus.\\
    $\mathcal{B}$                       & $\{ p_0, p_1, \cdots, p_{\alpha-1}\} $, a set of prime limbs of $P = \prod_{i=0}^{\alpha-1} p_i $.\\
    $\mathcal{C}$                       & $\{ q_0, q_1, \cdots, q_L\}$, a set of prime limbs of $Q = \prod_{i=0}^{L} q_i $.\\
    $\mathcal{C}_i$                     & Partial limb group of $\mathcal{C}$. $ \mathcal{C}_i = \{q_{\alpha i}, q_{\alpha i + 1}, \cdots, q_{\alpha (i + 1) - 1}\}$, $i = 0, 1, \cdots, \mathtt{dnum}\! -\! 1 $.\\
    $\mathcal{D}$                       & $\mathcal{B} \cup \mathcal{C}$, a set of prime limbs of $PQ$.\\
    $L$                                 & Maximum multiplicative level.\\
    $\ell$                              & (Current) multiplicative level. A level-$\ell$ polynomial has $ \ell + 1 $ limbs. $ 0 \leq \ell \leq L $.\\
    $\mathtt{dnum}$                     & Decomposition number.\\
    $\alpha$                            & $ (L + 1) / \mathtt{dnum} $. The number of prime limbs in $\mathcal{C}_i$ and $\mathcal{B}$.\\
    $ \Delta $                          & Scale multiplied during encryption.\\
    $ \mathbf{m} $                      & Message, a vector of $n$ slots.\\
    $\mathtt{P}, \mathtt{P}_\mathbf{m}$ & Polynomial. Subscript $ \mathbf{m} $ emphasizes that it corresponds to a message $\mathbf{m}$.\\
    $[\![ \mathbf{m} ]\!]$              & Ciphertext encrypting a message $ \mathbf{m} $.\\
    $[\mathtt{P}]_{q_i}$                & $q_i$-limb of $\mathtt{P}$.\\
    $[\mathtt{P}]_{\mathcal{C}}$        & $\{ [\mathtt{P}]_{q_i} : q_i \in {\mathcal{C}} \}$, the set of $q_i$-limbs of $\mathtt{P}$ for $q_i \in {\mathcal{C}}$.\\
    $\textbf{evk}$                      & Evaluation key. An $\textbf{evk}$ for HMult ($\textbf{evk}_{\text{mult}}$) and $\textbf{evk}$s for HRots with different rotation amount $r$'s ($\textbf{evk}_{\text{rot}}^{(r)}$'s) exist.\\
    \bottomrule
   \end{tabularx}
\end{table}
\renewcommand{\arraystretch}{1.0}
\setlength{\tabcolsep}{6pt}

\renewcommand{\arraystretch}{1.2}
\setlength{\tabcolsep}{3.8pt}
\begin{table*}[t]
  \caption{Primitive HE ops of CKKS}
  \label{table:operations}
  \centering
  \begin{tabularx}{\textwidth}{lXp{3.3in}}
    \toprule
    \textbf{Operation} & \textbf{Result} & \textbf{Description}\\
    \midrule
    $\text{CAdd}([\![\mathbf{m}]\!],c)$ & $[\![\mathbf{m}+\mathbf{c}]\!]=(\mathtt{B}_{\textbf{m}}+\textbf{c},\mathtt{A}_{\textbf{m}})$ & Add a scalar $c$ to a ciphertext. $\textbf{c}$ is a length-$N$ vector with every entry $c$.\\
    $\text{CMult}([\![\mathbf{m}]\!],c)$ & $[\![\mathbf{m} \cdot \mathbf{c}]\!]=(\mathtt{B}_{\textbf{m}} \cdot \textbf{c},\mathtt{A}_{\textbf{m}} \cdot \textbf{c})$ & Multiply a scalar to a ciphertext.\\
    $\text{PAdd}([\![\mathbf{m}]\!], \mathtt{P}_{\textbf{m}^\prime})$ & $[\![\mathbf{m}+\mathbf{m}^\prime]\!] = (\mathtt{B}_{\textbf{m}} + \mathtt{P}_{\textbf{m}^\prime},\mathtt{A}_{\textbf{m}})$ & Add an unencrypted polynomial to a ciphertext.\\
    $\text{PMult}([\![\mathbf{m}]\!], \mathtt{P}_{\textbf{m}^\prime})$ & $[\![\mathbf{m} \cdot \mathbf{m}^\prime]\!] = (\mathtt{B}_{\mathbf{m}} * \mathtt{P}_{\mathbf{m}^\prime},\mathtt{A}_{\mathbf{m}} * \mathtt{P}_{\mathbf{m}^\prime})$ & Multiply an unencrypted polynomial to a ciphertext.\\
    $\text{HAdd}([\![\mathbf{m}]\!],[\![\mathbf{m}^\prime]\!])$ & $[\![\mathbf{m}+\mathbf{m}^\prime]\!] = (\mathtt{B}_{\mathbf{m}} + \mathtt{B}_{\mathbf{m}^\prime},\mathtt{A}_{\mathbf{m}} + \mathtt{A}_{\mathbf{m}^\prime})$ & Add two ciphertexts.\\
    $\text{HMult}([\![\mathbf{m}]\!], [\![\mathbf{m}^\prime]\!], \mathbf{evk}_{\text{mult}})$ & $[\![\mathbf{m} \cdot \mathbf{m}^\prime]\!] = \text{KeySwitch}(\mathtt{A}_{\mathbf{m}} * \mathtt{A}_{\mathbf{m}^\prime}, \mathbf{evk}_{\text{mult}}) \ + $ & Multiply two ciphertexts.\\
    & $(\mathtt{B}_{\mathbf{m}} * \mathtt{B}_{\mathbf{m}^\prime}, \mathtt{A}_{\mathbf{m}} * \mathtt{B}_{\mathbf{m}^\prime} + \mathtt{A}_{\mathbf{m}^\prime} * \mathtt{B}_{\mathbf{m}})$ & \\
    $\text{HRot}([\![\mathbf{m}]\!],r, \mathbf{evk}_{\text{rot}}^{(r)})$ & $[\![\mathbf{m} \ll r]\!] = \text{KeySwitch}(\psi_r(\mathtt{A}_{\mathbf{m}}), \mathbf{evk}_{\text{rot}}^{(r)}) \ + $ & Perform a circular left shift by $r$ slots. $\psi_r$ is an automorphism performed\\
    & $(\psi_r(\mathtt{B}_{\mathbf{m}}), \mathbf{0})$ & on polynomials.\\
    $\text{HRescale}([\![\mathbf{m}]\!])$ & $[\![\Delta^{-1} \cdot \mathbf{m}]\!] = (\Delta^{-1}\mathtt{B}_{\mathbf{m}}, \Delta^{-1}\mathtt{A}_{\mathbf{m}})$ & Restore the scale of a ciphertext having scale $\Delta^2$ back to $\Delta$.\\
    \bottomrule
  \end{tabularx}
\end{table*}
\renewcommand{\arraystretch}{1.0}
\setlength{\tabcolsep}{6pt}

\subsection{Computational optimizations for HE ops}
\label{sec:2_2_ntt_and_rns}

HE schemes use a fixed set of computations called \emph{HE ops} to process encrypted data without decryption.
However, HE ops are expensive.
A polynomial in $\mathrm{R}_Q$ is a long ($N=2^{16}$) vector of large integers ($Q > 2^{1000}$).
In general, two techniques are used to mitigate the cost of HE ops.

\noindent\textbf{Number-Theoretic Transform (NTT):} NTT is used to lower the complexity of a polynomial multiplication (mult).
Multiplying two polynomials in $\mathrm{R}_Q$ is a (negacyclic) convolution of two length-$N$ vectors and has an $\mathcal{O}(N^2)$ complexity.
NTT is a variant of DFT, which converts the convolution into an element-wise mult, reducing the complexity to $\mathcal{O}(N)$:
\begin{equation*} \label{eq:ntt}
    \text{NTT}(\mathtt{P}_1 * \mathtt{P}_2) = \text{NTT}(\mathtt{P}_1) \cdot \text{NTT}(\mathtt{P}_2) 
\end{equation*}
While NTT reduces the complexity of a polynomial mult, there is an additional cost of performing NTT.
Thus, Fast Fourier Transform (FFT) algorithms~\cite{cooley-tukey, gentleman-sande} with $\mathcal{O}(N \log N)$ complexities are applied to compute NTT, and we keep polynomials in their NTT-applied versions to prevent unnecessary NTTs and Inverse NTTs (INTTs).
We refer to this NTT-applied version as the \emph{evaluation representation} of the polynomial, and we assume every polynomial is in its evaluation representation unless otherwise stated.
However, as there are functions that cannot be performed when a polynomial is in its evaluation representation, we often perform INTT to bring back the polynomial to the original representation, which we refer to as the \emph{coefficient representation}.
 
\noindent\textbf{Residue Number System (RNS) and limbs:} $\mathtt{P}_{\mathbf{m}}$ has coefficients in $ \mathbb{Z}_Q $ having up to 1,000s of bits.
RNS is widely used to reduce the high complexity of performing multi-precision arithmetic between large integer coefficients~\cite{sac-2018-frns-ckks, sac-2016-behz}.
We first choose $(L+1)$ word-sized prime numbers to form an ordered set $\mathcal{C}=\{q_0, \cdots , q_L\}$ such that $ Q=\prod_{i=0}^{L}q_i $; we refer to each $q_i$ as a \emph{prime limb} of $Q$.
Based on the Chinese Remainder Theorem (CRT), any coefficient $c \in \mathbb{Z}_Q $ can be uniquely represented with a vector $(c \text{ mod } q_0, \cdots, c \text{ mod } q_L)$, allowing a single polynomial $\mathtt{P}_{\mathbf{m}} \in \mathrm{R}_Q$ to be decomposed into $(L+1)$ polynomials in $ \{ \mathrm{R}_{q_i}\}_{0 \leq i \leq L } $.
Under such representation, we treat $\mathtt{P}_{\mathbf{m}}$ as a matrix of size $(L+1)\times N$ words and denote it as $[\mathtt{P}_{\mathbf{m}}]_\mathcal{C}$.
We refer to each row of the matrix as the polynomial's \emph{limb}. 
We use the term $q_i$\emph{-limb} to indicate a specific row, for example, the $i$-th row.
It is denoted as $[\mathtt{P}_{\mathbf{m}}]_{q_i}$ and is obtained by applying modulo $q_i$ to the coefficients.
Using RNS and NTT, a polynomial op becomes element-wise ops between $(L+1)\times N$ matrices that only involve word-sized arithmetic.

\begin{figure}[t]
\vspace{-1.0\intextsep}
\begin{algorithm}[H]
\caption{BConvRoutine}\label{alg:bconvroutine}
\algnewcommand\algorithmicnotation{\textbf{Notation:}}
\algnewcommand\Notation{\item[\algorithmicnotation]}
\begin{algorithmic}[1]
\Notation{$\mathtt{P}_\mathtt{coeff}$ = coefficient representation of $\mathtt{P}$}
\Procedure{BConvRoutine}{$[\mathtt{P}]_\mathcal{B}, \mathcal{C}$}
    \State $[\mathtt{P}_\mathtt{coeff}]_\mathcal{B} \leftarrow \text{INTT}([\mathtt{P}]_\mathcal{B})$
    \State $[\mathtt{P}_\mathtt{coeff}]_\mathcal{C} \leftarrow \underset{\mathcal{B} \to \mathcal{C}}{\textrm{BConv}}([\mathtt{P}_\mathtt{coeff}]_\mathcal{B})$
    \State $[\mathtt{P}]_\mathcal{C} \leftarrow \text{NTT}([\mathtt{P}_\mathtt{coeff}]_\mathcal{C})$ 
    \State \textbf{return} $[\mathtt{P}]_\mathcal{C}$
\EndProcedure
\end{algorithmic}
\end{algorithm}
\vspace{-1.0\intextsep}
\end{figure}

Base conversion (BConv)~\cite{sac-2016-behz} is used to change the prime limb set of a polynomial.
To perform BConv, the polynomial must be in its coefficient representation.
Thus, it is common in CKKS to perform a series of INTT $\rightarrow$ BConv $\rightarrow$ NTT, which we refer to as a \emph{BConvRoutine} (Alg.~\ref{alg:bconvroutine}). 
BConv converts $[\mathtt{P}_{\mathtt{coeff}}]_{\mathcal{B}}$ for a prime limb set $\mathcal{B} =\{p_0, p_1, \cdots, p_{\alpha-1}\}$ into $[\mathtt{P}_{\mathtt{coeff}}]_{\mathcal{C}}$ by Eq.~\ref{eq:bconv}, where
$\hat{p}_=\prod_{i \neq j} p_i$ for $\forall p_i \in \mathcal{B}$.
\begin{equation} \label{eq:bconv}
\begin{split}
    &[\mathtt{P}_{\mathtt{coeff}}]_{\mathcal{C}} = \underset{\mathcal{B} \rightarrow \mathcal{C}}{\text{BConv}}([\mathtt{P}_{\mathtt{coeff}}]_{\mathcal{B}})\\
    &= \bigg\{ \sum_{j=0}^{\alpha\!-\!1}\! \big( [\mathtt{P}_\mathtt{coeff}]_{p_j} \! \cdot \hat{p}_j^{-1} \text{ mod }p_j\big) \cdot \hat{p}_j \text{ mod } {q_i}\bigg\}_{0 \leq i \leq L}
\end{split}
\end{equation}
We further explain the details of BConv in Section~\ref{sec:5_2_bconvu1}.

\subsection{Primitive HE ops in CKKS}
\label{sec:2_3_he_ops}

In Table~\ref{table:operations}, we summarize the \emph{primitive} HE ops of CKKS.
These HE ops become the building blocks of complex HE ops such as homomorphic linear transform.
We describe some of the important details of the primitive HE ops below.

\noindent\textbf{HRescale:} After multiplying (CMult, PMult, or HMult) a ciphertext with another operand having a scale of $\Delta$, the scale of the result becomes ${\Delta}^2$.
HRescale is a primitive HE op that divides the ciphertext by $\Delta$ to recover its scale back to $\Delta$.

As it is difficult to perform a division when polynomials are represented in limbs using RNS, CKKS utilizes an alternative way to obtain the divided result.
In CKKS, the value of each $q_i$ is set close to $\Delta$ and HRescale is performed by eliminating the last limb of the polynomial ($q_L$-limb) and multiplying $q_L^{-1} \text{ mod } q_i (i\!=\!0,1, \cdots, L-1)$ to the remaining limbs~\cite{sac-2018-frns-ckks}.
We can successively apply HRescale after each mult until the polynomial only has the $q_0$-limb left.
At this moment, it is impossible to perform additional mults.
Thus, the maximum number of possible mults is $L$, which we refer to as the \emph{maximum multiplicative level}.
We call the remaining number of possible mults the \emph{(current) multiplicative level} and denote it as $\ell$.
A ciphertext at a multiplicative level $\ell$ is represented as a pair of $(\ell + 1)\times N$ matrices.

\noindent\textbf{Automorphism:} To circularly shift a message by $r$ slots through HRot, we first compute the automorphism of polynomials.
Automorphism moves the $i$-th coefficient of a polynomial to the position of the $\psi_r(i)$-th coefficient by applying the mapping in Eq. \ref{eq:automorphism} to each limb.
We reuse the notation to denote automorphism on a polynomial $\mathtt{P}_{\textbf{m}} $ as $\psi _r(\mathtt{P}_{\textbf{m}})$ and refer to $r$ as the \emph{rotation amount}.
\begin{equation} \label{eq:automorphism}
    \psi_r : i\mapsto i \cdot 5^r \text{ mod } N
\end{equation}

\noindent\textbf{Key-switching:} HMult and HRot produce polynomials that are only decryptable with keys $\mathtt{S}^2$ or $\psi_r(\mathtt{S})$.
To make such a polynomial decryptable with the secret key $\mathtt{S}$, 
we perform \emph{key-switching} during HMult or HRot by multiplying it with a special public key called the \emph{evaluation key} ($\textbf{evk}$):
\begin{equation}
    \text{KeySwitch}(\mathtt{P}_\mathbf{m}, \mathbf{evk}) = P^{-1} \cdot \mathtt{P}_\mathbf{m} \boldsymbol{\cdot} \mathbf{evk}
\end{equation}
In particular, the $\textbf{evk}$ encrypting $\mathtt{S}^2$ is used for HMult and is denoted as $\textbf{evk}_{\text{mult}}$.
The $\textbf{evk}$ encrypting $\psi_r(\mathtt{S})$ is used for HRot and is denoted as $\textbf{evk}_{\text{rot}}^{(r)}$.
A separate $\textbf{evk}_{\text{rot}}^{(r)}$ is required for each rotation amount $r$.

In general, key-switching is an expensive HE op as it involves multiple NTTs and BConvs.
Moreover, because key-switching is performed frequently, it dominates the execution time of an HE application~\cite{tches-2021-100x}.
Multiplying $ \mathtt{P}_{\mathbf{m}} $ and $ \mathbf{evk} $ requires additional steps as they have different modulus.
While $ \mathtt{P}_{\mathbf{m}} \in \mathrm{R}_Q$ has modulus $Q$, $ \mathbf{evk} \in  \mathrm{R}_{PQ}$ has modulus $PQ$ where $P$ is the \emph{special modulus} that can be decomposed into a set of $ \alpha $ prime limbs $\mathcal{B}=\{ p_0, p_1, \cdots, p_{\alpha-1}\}$.
BConvs are performed to bring $\mathtt{P}_{\mathbf{m}}$ into $\mathrm{R}_{PQ}$ by generating the $[\mathtt{P}_{\mathbf{m}}]_\mathcal{B}$ part.
After multiplying $\mathtt{P}_{\mathbf{m}}$ with $ \mathbf{evk}$, BConvs bring the multiplied results back to $\mathrm{R}_Q$, and the results are divided by $P$ to reduce error in a similar way to HRescale.

\begin{figure}[t]
\vspace{-1.0\intextsep}
\begin{algorithm}[H]
\caption{Generalized key-switching}\label{alg:key-switching}
\algnewcommand\algorithmicnotation{\textbf{Notation:}}
\algnewcommand\Notation{\item[\algorithmicnotation]}
\begin{algorithmic}[1]
\Notation{$\mathbf{evk} = \{\mathbf{evk}_i\}_{0 \le i < \mathtt{dnum}}, \mathtt{P} = \{[\mathtt{P}]_{\mathcal{C}_i}\}_{0 \le i < \mathtt{dnum}} $}
\Procedure{KeySwitch}{$ \mathtt{P}$, $ \mathbf{evk} $}
  \For{i $\leftarrow$ 0, 1,$\cdots$, $\mathtt{dnum}$-1} \label{alg:key-switching:line:for}
    \State $ [\mathtt{P}_{i}]_\mathcal{D} \leftarrow [\mathtt{P}]_{\mathcal{C}_i} \cup \text{BConvRoutine}([\mathtt{P}]_{\mathcal{C}_i}, \mathcal{D}\setminus \mathcal{C}_i) $ \label{alg:key-switching:line:bconvroutine}
  \EndFor
  \State $(\mathtt{B}, \mathtt{A}) \leftarrow \sum_{i=0}^{\mathtt{dnum}-1} \text{PMult}(\mathbf{evk}_i, [\mathtt{P}_{i}]_\mathcal{D}) $ \label{alg:key-switching:line:accum}
  \State $ \mathtt{B}^{\prime} \leftarrow [\mathtt{B}]_\mathcal{C} - \text{BConvRoutine}([\mathtt{B}]_\mathcal{B}, \mathcal{C})$
  \label{alg:key-switching:line:bconvroutine:B}
  \State $ \mathtt{A}^{\prime} \leftarrow [\mathtt{A}]_\mathcal{C} - \text{BConvRoutine}([\mathtt{A}]_\mathcal{B}, \mathcal{C})$
  \label{alg:key-switching:line:bconvroutine:A}
  \State \textbf{return} $(P^{-1} \cdot \mathtt{B}^{\prime}, P^{-1} \cdot \mathtt{A}^{\prime})$
  \label{alg:key-switching:line:bconvroutine:P-1}
\EndProcedure
\end{algorithmic}
\end{algorithm}
\vspace{-1.9\intextsep}
\end{figure}

\noindent\textbf{Security of HE and generalized key-switching:} 
We use the generalized key-switching method in \cite{rsa-2020-better}.
It introduces a new parameter, $ \mathtt{dnum} $, and sets the number of special prime limbs as
$ \alpha = (L + 1) / \mathtt{dnum} $.
A single $ \mathbf{evk} $ consists of $ \mathtt{dnum} $ pairs of polynomials: $\mathbf{evk} =  ( \mathbf{evk}_0,  \mathbf{evk}_1, \cdots, \mathbf{evk}_{\mathtt{dnum} - 1} )$, $\mathbf{evk}_i \in \mathrm{R}_{PQ}^2$.
Also, the prime limb set $\mathcal{C}$ is decomposed into $\mathtt{dnum}$ \emph{partial limb groups}: $ \mathcal{C}_i = \{ q_{\alpha i}, q_{\alpha i + 1}, \cdots,  q_{\alpha (i + 1) - 1}\}$ ($0 \leq i < \mathtt{dnum}$).
$(L+1)$ limbs of the polynomial $\mathtt{P}$ are decomposed into $ \mathtt{dnum} $ pieces as well: $\{ [\mathtt{P}]_{\mathcal{C}_i} \}_{0 \leq i < \mathtt{dnum}}$.
Before multiplying each $\mathbf{evk}_i$ with the corresponding piece $[\mathtt{P}]_{\mathcal{C}_i}$, we perform BConvRoutine on each piece $[\mathtt{P}]_{\mathcal{C}_i}$ to extend the limbs to $\mathrm{R}_{PQ}$ (line~\ref{alg:key-switching:line:bconvroutine} in Alg.~\ref{alg:key-switching}).
Then, we multiply $\mathtt{dnum}$ pairs of $\mathbf{evk}_i$s each with $[\mathtt{P}_i]_{\mathcal{D}}$ and accumulate them (line~\ref{alg:key-switching:line:accum}).
BConvRoutines bring the result back to $\mathrm{R}_{Q}$ (line~\ref{alg:key-switching:line:bconvroutine:B}-\ref{alg:key-switching:line:bconvroutine:P-1}).

The computational complexity of key-switching and the size of an $ \mathbf{evk} $ increase as $ \mathtt{dnum} $ gets larger, degrading overall performance.
Despite such limitations, generalized key-switching is used to acquire more multiplicative levels.
The \emph{security level} of HE is roughly decided by $N$ and $( \alpha + L + 1) $, which increases as $N$ gets larger or $( \alpha + L + 1) $ gets smaller~\cite{wahc-2019-curtis}.
Recent HE studies~\cite{eurocrypt-2021-efficient,eurocrypt-2022-varmin,eurocrypt-2021-invsine} and libraries~\cite{github-lattigo, online-palisade} typically target a security level of 128 bits.
Under such security level constraints, $N$ is chosen as low as possible and $( \alpha + L + 1) $ as high as possible.
Table~\ref{tab:params} shows exemplar parameter selections satisfying 128-bit security.
If $\mathtt{dnum}$ is large, the value of $ \alpha = (L + 1) / \mathtt{dnum} $ reduces, enabling a higher $L$ under the same  $( \alpha + L + 1)$.
In particular, when using a small $N$, it is inevitable to use a $\mathtt{dnum}$ value close to its maximum ($=L+1$) to secure as many multiplicative levels as possible.

\subsection{CKKS bootstrapping}
\label{sec:2_4_bootstrap}

When the ciphertext $[\![ \mathbf{m} ]\!] $'s multiplicative level $\ell$ reaches 0 after multiple HRescales, it becomes impossible to perform more mult ops.
As such, one must recover the multiplicative level through a special op called \emph{bootstrapping} to enable further ops on $[\![ \mathbf{m} ]\!]$.
Because bootstrapping allows an unlimited number of ops, it becomes an essential op for complex HE applications.
The HE schemes that support bootstrapping are generally referred to as \emph{Fully HE} (FHE).

Bootstrapping is a complex sequence of HE ops that consists of four major steps: LevelRecover, Homomorphic IDFT (H-IDFT), EvalMod, and Homomorphic DFT (H-DFT).
During LevelRecover, we raise $[\![ \mathbf{m} ]\!] $'s multiplicative level to its maximum, $L$, by changing the modulus from $q_0$ to $Q$.
However, if we decrypt it using the secret key $\mathtt{S}$, we get $\mathtt{P}_\mathbf{m'}$ instead of $\mathtt{P}_\mathbf{m}$ due to the multiple of $q_0$, which was previously removed through modular reduction (i.e., $\mathtt{P}_{\mathbf{m'}} = \mathtt{P}_{\mathbf{m}} + q_0 \cdot \mathtt{I} $).
To remove the $q_0 \cdot \mathtt{I}$ term from $\mathtt{P}_{\mathbf{m'}}$, we first apply H-IDFT on $[\![ \mathbf{m'} ]\!]$ to enable homomorphic ops on $\mathtt{P}_{\mathbf{m'}}$ instead of $\mathbf{m'}$.
Based on Eq.~\ref{eq:encode}, we denote the resulting $[\![\text{IDFT}(\mathbf{m'})]\!]\!$ as $\![\![ \mathtt{P}_{\mathbf{m'}} ]\!]$ to emphasize that the ciphertext encrypts the coefficients of $\mathtt{P}_{\mathbf{m'}}$.
EvalMod homomorphically performs the modulo operation on $ [\![ \mathtt{P}_{\mathbf{m'}} ]\!] $.
The modulo operation is not a linear function, so it is approximated by a high-degree polynomial function.
Homomorphic evaluation of the function consists of multiple HMults and CMults, consuming many multiplicative levels~\cite{eurocrypt-2018-heaanboot}.
This step recovers $ [\![ \mathtt{P}_{\mathbf{m}} ]\!] $ by calculating $ [\![ \mathtt{P}_{\mathbf{m'}}\text{ mod }q_0 ]\!] $ and H-DFT recovers $ [\![ \mathbf{m}]\!]$ from $ [\![ \mathtt{P}_{\mathbf{m}} ]\!] $.

The resulting ciphertext has $ (L - L_{boot}) $ multiplicative levels as multiple HRescales consume $ L_{boot} $ levels during bootstrapping.
Thus, $L$ must be high enough to support CKKS bootstrapping.
However, the security limit requires that $L$ can only be increased proportionally to $N$, so $N$ should be sufficiently high ($\geq 2^{15}$) to allow bootstrapping~\cite{eurocrypt-2022-varmin}.
Prior works on FHE~\cite{github-lattigo,tches-2021-100x,eurocrypt-2022-varmin, eurocrypt-2021-efficient} mostly use $N = 2^{15}$ to $2^{17}$ to preserve sufficient  multiplicative levels for other ops.
Thus, we use the parameter set summarized in Table \ref{tab:params} in this work, which is a slightly modified version of Lattigo~\cite{github-lattigo}'s parameter set and guarantees 128-bit security.

\setlength{\tabcolsep}{5pt}
\renewcommand{\arraystretch}{1.2}
\begin{table}[t]
\caption{Representative parameters used in HE acceleration works and corresponding data sizes}
\label{tab:params}
\centering
\begin{threeparttable}
\begin{tabular}{l|ccccc|ccc}
\hline
\multirow{2}{*}{Work}                             & \multicolumn{5}{c|}{Parameters}                                 & \multicolumn{3}{c}{Data size (MB)} \\
                                                  & $N$      & $L$ & $L_{\text{boot}}$ & $\mathtt{dnum}$ & $\alpha$ & $\mathtt{P}_{\mathbf{m}}$ & $[\![\mathbf{m}]\!]$ & $\mathbf{evk}$ \\
\hline
\hline
Lattigo~\cite{github-lattigo}                     & $2^{16}$ & 24  & 15                & 5               & 5        & 12.5                      & 25                   & 150 \\
100x~\cite{tches-2021-100x}\textsuperscript{\dag} & $2^{17}$ & 29  & 19                & 3               & 10       & 30                        & 60                   & 240 \\
F1~\cite{micro-2021-f1}                           & $2^{14}$ & 15  & -                 & 16              & 1        & 1                         & 2                    & 34  \\
\NAME                                             & $2^{16}$ & 23  & 15                & 4               & 6        & 12                        & 24                   & 120 \\
\hline
\end{tabular}
\begin{tablenotes}
  \item[\dag] \cite{tches-2021-100x} used a 173-bit secure parameter set, while others 128-bit.
\end{tablenotes}
\end{threeparttable}
\end{table}
\setlength{\tabcolsep}{6pt}
\renewcommand{\arraystretch}{1.0}

\section{FHE Memory Bottlenecks}
\label{sec:3_bottleneck}

\subsection{Limitations of prior HE accelerators}
\label{sec:3_1_prior_accelerators}

All HE ops can be broken down into \emph{primary functions} that can be executed in parallel: (I)NTT, BConv, automorphism, and other \emph{element-wise functions}.
By using well-known FFT algorithms~\cite{cooley-tukey}, (I)NTT can be effectively parallelized by dividing it into $\log N$ stages, where each stage consists of parallel $\sfrac{N}{2}$ butterfly ops.
Since $N$ is as large as $2^{16}$, the degree of parallelism is high.
Given such, domain-specific architectures that employ a massive number of functional units (FUs) to reap such opportunities become an appealing option for accelerating HE.
To this end, designing a memory system capable of feeding these abundant FUs on-chip becomes a primary design objective.

Prior work has identified that HE ops have very low arithmetic intensity and are bounded by the off-chip memory bandwidth~\cite{arxiv-2021-does, tches-2021-100x, hpca-2019-roy}.
Even a simple HE op, such as HMult, has a working set including several ciphertexts and an \evk, which amounts to several hundreds of MBs.
Unfortunately, conventional CPU/GPU/FPGA solutions typically do not come with on-chip memory large enough to hold the entire working set of HE ops (e.g., even the latest NVIDIA H100 GPU ``only'' has 50MB of on-chip cache~\cite{cse-2022-h100}), causing frequent off-chip memory accesses and high memory bandwidth pressure.

F1~\cite{micro-2021-f1} is the first ASIC HE accelerator to evaluate the performance of a single-slot CKKS bootstrapping.
F1 uses small parameters, $(N, L) = (2^{14}, 15)$, and deploys a massive amount of computation logic to utilize the parallelism in HE ops.
In particular, F1 utilizes a dedicated \emph{NTT unit} (NTTU) that implements a pipelined 2D-FFT using  $\frac{1}{2}\sqrt{N} \log N\! =\! 896$ modular multipliers.
F1 consists of 16 vector \emph{clusters}, each with $\sqrt{N}\! =\! 128$ lanes that feed data to an NTTU and other FUs.
Overall, F1 has a massive amount of 18,432 modular multipliers in a chip: 14,336 for NTTUs and 4,096 for element-wise multipliers.
Every HE op is essentially a sequence of modular mults and modular adds, and a modular mult requires a costly modular reduction op~\cite{eurocrypt-1986-barrett, 1985-montgomery}.
Thus, the computational capability of an HE accelerator can be quantified by the number of modular multipliers.

Although F1 provides high speedup compared to conventional systems, its applicability is limited to small problem sizes and simple workloads; when it comes to practical FHE parameters and complex workloads that require frequent bootstrapping, the performance of F1 can be degraded significantly.\footnote{Because of the small parameters used in F1, F1 cannot perform bootstrapping of ciphertexts encrypting a message vector, only supporting a message with a single slot ($n=1$).
Thus, F1's bootstrapping throughput is severely limited.
Single-slot bootstrapping does not need H-(I)DFT steps, so it is less costly and consumes fewer levels.
F1 also uses 32-bit machine word and prime sizes, but practical bootstrapping requires the use of larger primes for error mitigation.}
We first explain the details of bootstrapping for analysis.

\subsection{State-of-the-art bootstrapping algorithm}
\label{sec:3_2_sota_bootstrapping}

\begin{figure}[t]
\vspace{-1.0\intextsep}
\begin{algorithm}[H]
\caption{FFT-like homomorphic DFT (Radix-$2^k$)}\label{alg:fft-like-hdft}
\begin{algorithmic}[1]
\Require{Precomputed plaintexts (polynomials) $\{\mathtt{P}_{s,i}\}$}
\Procedure{DFT}{$[\![\mathbf{m}]\!], \{\mathtt{P}_{s,i}\}$}
\State{$[\![\mathbf{t}]\!] \leftarrow [\![\mathbf{m}]\!]$}
\For{$s \leftarrow$ 0, 1, $\cdots$, $\log_{2^k}\!n-1$\label{alg:fft-like-hdft:line:for}}
  \State $[\![\mathbf{t}]\!] \leftarrow  \sum_{i=-2^k\!+\!1}^{2^k\!-\!1} \text{PMult}(\text{HRot}([\![\mathbf{t}]\!], i\! \cdot\! 2^{k \cdot s}), \mathtt{P}_{s,i})$\label{alg:fft-like-hdft:line:sum}
  \State $[\![\mathbf{t}]\!] \leftarrow \text{HRescale}([\![\mathbf{t}]\!])$ \Comment{consumes one level}
\EndFor
\State{\Return{$[\![\mathbf{t}]\!]$}}
\EndProcedure
\end{algorithmic}
\end{algorithm}
\vspace{-1.0\intextsep}
\end{figure}

H-(I)DFT is a memory-intensive op that takes up most of the execution time in bootstrapping.
H-(I)DFT can be performed in a similar pattern to FFT as shown in Alg.~\ref{alg:fft-like-hdft}, where $2^k$ denotes the radix in FFT~\cite{cooley-tukey}.
Each iteration requires ($2^{k+1}-1$) HRots and ($2^{k+1}-1$) PMults, each with a different $\mathbf{evk}_{\text{rot}}^{(r)}$ and a different \emph{plaintext}, a polynomial for precomputed DFT constants ($\mathtt{P}_{s,i}$).
The whole process consumes $\log_{2^k}\!n$ multiplicative levels, so the choice of $k$ creates a trade-off between computation and level consumption.

Line~\ref{alg:fft-like-hdft:line:sum} of Alg.~\ref{alg:fft-like-hdft} can be further improved by first performing the \emph{pre-rotation} in Eq.~\ref{eq:prerotation},
\begin{equation} \label{eq:prerotation}
[\![\mathbf{t}^\prime]\!] = \text{HRot}([\![\mathbf{t}]\!], -2^k\cdot2^{k \cdot s})
\end{equation}
then applying the BSGS (Baby-Step, Giant-Step) algorithm~\cite{eurocrypt-2018-heaanboot} as shown in Eq.~\ref{eq:bsgs}. $k_1$ and $k_2$ in Eq.~\ref{eq:bsgs} are integers satisfying $k+1 = k_1 + k_2$.
Consequently, the number of HRots reduces from $\mathcal{O}(2^k)$ to $\mathcal{O}(\sqrt{2^k})$.
\begin{equation} \label{eq:bsgs}
\underbrace{\sum_{\mathclap{j=0}}^{\mathclap{2^{k_2}\!-1}}  \text{HRot}  \big(  \sum_{\mathclap{i=0}}^{\mathclap{2^{k_1}\!-1}}  \text{PMult}(\underbrace{\text{HRot}([\![\mathbf{t}^\prime]\!],i\! \cdot\! 2^{k\cdot s})}_{\text{baby-step}}, \mathtt{P}_{s,i,j}^\prime), j\!\cdot\!2^{k_1\! + k\cdot s}  \big)}_{\text{giant-step}}
\end{equation}

In \NAME with $n = 2^{15}$, we use $k = 5$, which consumes 3 levels, and $(k_1, k_2)$ in Eq.~\ref{eq:bsgs} as $(3, 3)$.
With additional optimizations, 40 HRots and 158 PMults are performed for each H-(I)DFT, and
40 distinct $\textbf{evk}_{\text{rot}}^{(r)}$s and 158 plaintexts must be prepared, which amount to 6.4GB (resp., 0.6GB) of data for H-IDFT (H-DFT) (see Table~\ref{tab:params}).

For the entire bootstrapping algorithm, which we use as the baseline algorithm throughout this paper, we adopted the implementation from \cite{rsa-2020-better} while applying parts of the optimizations from \cite{eurocrypt-2018-slimboot, eurocrypt-2021-efficient, eurocrypt-2022-varmin}.
Refer to the papers for a more detailed explanation of the bootstrapping implementation.

\subsection{Memory bottleneck in bootstrapping}
\label{sec:3_3_off_chip_bottleneck}

The time to load $\mathbf{evk}_{\text{rot}}^{(r)}$s and plaintexts, which are used once per H-(I)DFT, becomes the hard bound of latency for H-(I)DFT.
Even if  we assume that integrating hundreds of MBs of on-chip memory is feasible with the current fabrication technology~\cite{youtube-dojo,hcs-2021-graphcore,hcs-2021-sambanova, isca-2020-groq}, we must fetch the single-use data of H-(I)DFT from off-chip memory due to its large aggregate size and limited data reuse opportunities.
With the latest HBM3~\cite{jedec-2022-hbm3}, a system having a 3TB/s off-chip memory bandwidth is feasible~\cite{cse-2022-h100}, which would allow loading the single-use data in 2.1ms (resp., 0.2ms) for H-IDFT (H-DFT).

We analyze the impact of such hard latency bound on domain-specific architectures by measuring the maximum utilization of FUs in F1.
We scaled F1 to the bootstrappable parameter we use (see Table~\ref{tab:params}) by utilizing NTTUs that feature $ \frac{1}{2}\sqrt{N} \log N\!=\!2,048$ modular multipliers to fully support bootstrapping.
As a result, the total number of modular multipliers in the chip increases by $2.22\times$ to 40,960.
We first computed the number of possible modular mults in the scaled F1 in 2.1ms (resp., 0.2ms) by assuming that it runs at 1GHz and is fully-pipelined so that 40,960 modular mults are possible every cycle.
Then, we divided it by the number of modular mults performed in H-IDFT (H-DFT) to derive the maximum achievable utilization of modular multipliers.
Overall, our analysis revealed that the scaled F1 performs poorly when handling off-chip memory bandwidth bottleneck caused by bootstrapping, only achieving an 8.61\% (13.32\%) utilization rate for H-IDFT (H-DFT).

Resolving the off-chip memory bandwidth bottleneck in bootstrapping should precede improving the computational capability for practical FHE workloads that require frequent bootstrapping, especially for domain-specific architectures that can utilize a massive amount of computational logic.  

\section{Algorithms tackling memory bottleneck}
\label{sec:4_algorithm}

With the advancement of fabrication technology, the scaling of off-die memory bandwidth is slower than that of logic density~\cite{isca-2009-disaggregated}.
We present our fabrication-technology-aware algorithmic enhancements that reduce off-chip memory access, alleviating FHE's memory bandwidth bottleneck.

\subsection{Minimum key-switching (Min-KS)}
\label{sec:4_1_multi_hop}

\begin{figure}[t]
  \hspace{0.05\columnwidth}\subfigure[Baseline BSGS\label{fig:BSGS}] {\includegraphics[width=0.9\columnwidth]{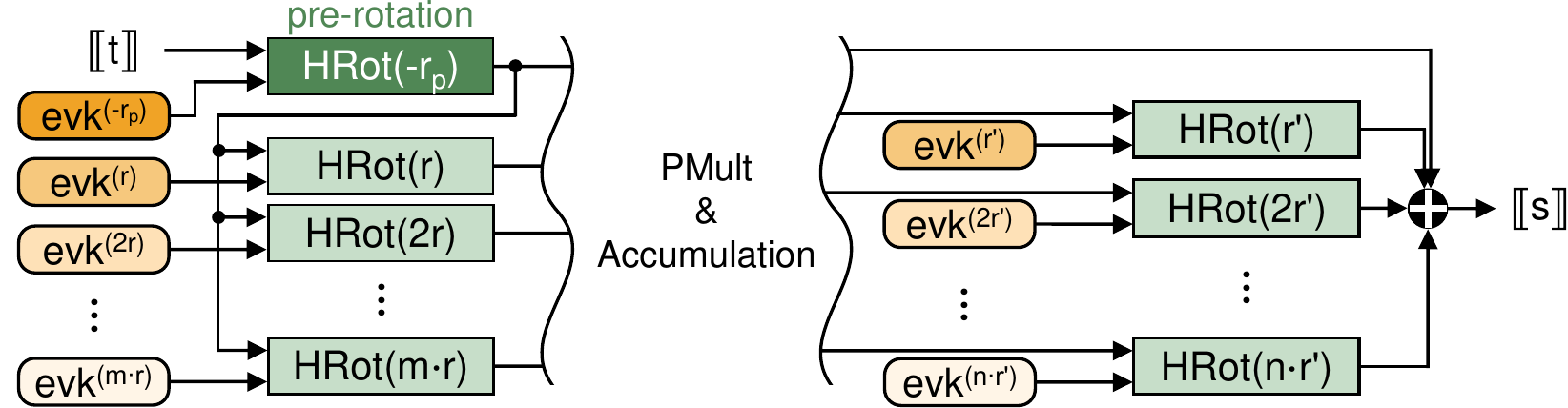}}
  \vskip0.0in
  \hspace{0.05\columnwidth}\subfigure[BSGS using minimal key-switching strategy in \cite{crypto-2018-linear}\label{fig:BSGS_minimal}] {\includegraphics[width=0.9\columnwidth]{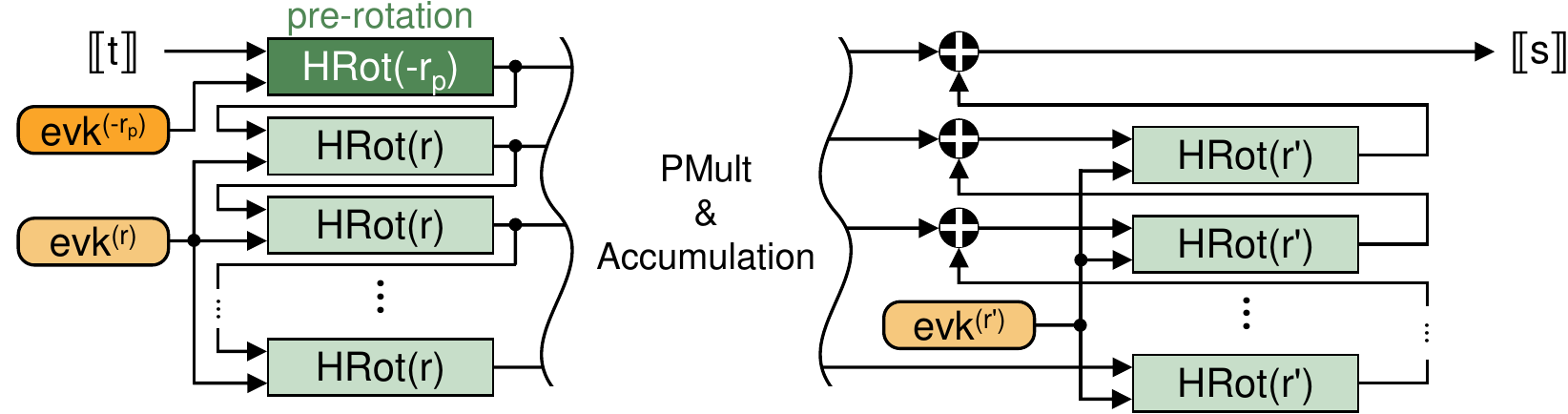}}
  \vskip0.0in
  \hspace{0.045\columnwidth}\subfigure[BSGS using our minimum key-switching (Min-KS)\label{fig:BSGS_multihop}] {\includegraphics[width=0.905\columnwidth]{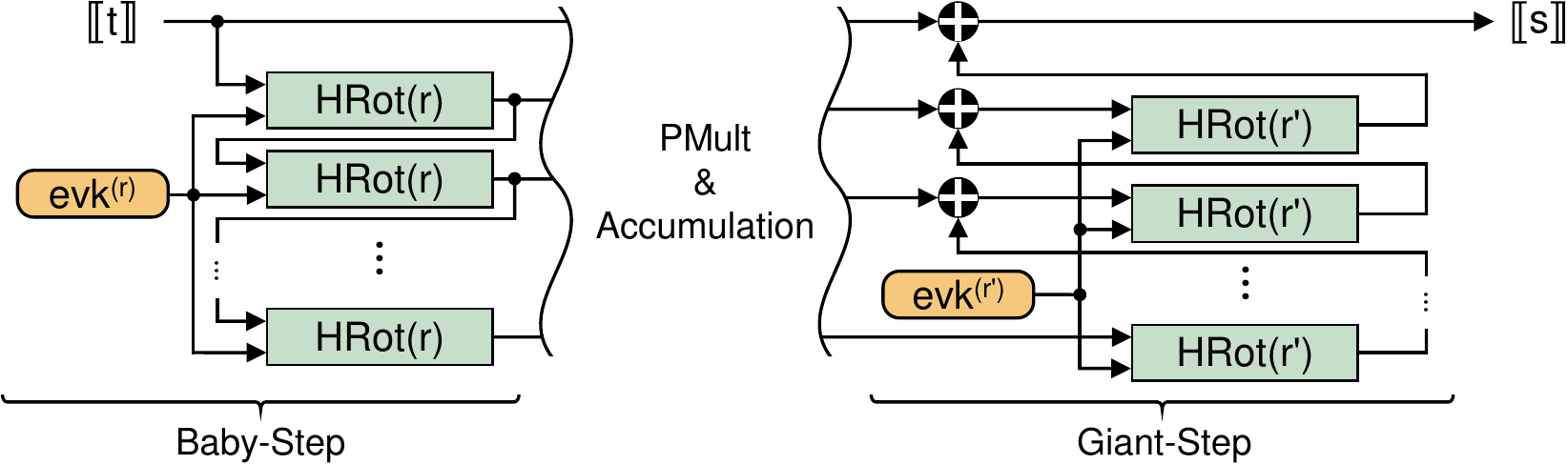}}
  \caption{Computational flow of the BSGS algorithm in H-(I)DFT (a) with the baseline algorithm (Eq.~\ref{eq:bsgs}), (b) with minimal key-switching strategy in \cite{crypto-2018-linear}, and (c) with our minimum key-switching (Min-KS).}
  \label{fig:multi-hop}
\end{figure}

\cite{crypto-2018-linear} proposed \emph{minimal key-switching} strategy for H-(I)DFT, which can be applied to the two following computational patterns found in H-(I)DFT, Eq.~\ref{eq:pattern1} corresponding to the baby-step and Eq.~\ref{eq:pattern2} to the giant-step.
\begin{equation} \label{eq:pattern1}
    \left\{\text{HRot}\left([\![\mathbf{x}]\!], i \cdot r, \textbf{evk}_{\text{rot}}^{(i \cdot r)}\right)\right\}_{0 \le i \le m}
\end{equation}
\begin{equation} \label{eq:pattern2}
    \sum_{i=0}^{m}\text{HRot}\left([\![\mathbf{x}_i]\!], i \cdot r, \textbf{evk}_{\text{rot}}^{(i \cdot r)}\right)
\end{equation}
Eq.~\ref{eq:pattern1} involves rotating the same ciphertext, and Eq.~\ref{eq:pattern2} involves rotating and accumulating different ciphertexts, each with rotation amounts showing an \emph{arithmetic progression} ($i\cdot r$).
Both computational patterns must load $m$ different \evks ($\mathbf{evk}_{\text{rot}}^{(r)}, \mathbf{evk}_{\text{rot}}^{(2r)}, \cdots, \mathbf{evk}_{\text{rot}}^{(m \cdot r)}$) from off-chip memory (see Fig.~\ref{fig:BSGS}), which makes HE workloads bottlenecked by the off-chip memory bandwidth.
\cite{crypto-2018-linear} instead computes each pattern iteratively, while reusing previous HRot computation results as in Eq.~\ref{eq:hrot-chain}.
Then, all HRots have the same rotation amount $r$, and consequently use the same single $\mathbf{evk}_{\text{rot}}^{(r)}$.
\begin{equation}\label{eq:hrot-chain}
\text{HRot}([\![\mathbf{x}]\!], i \cdot r)\! =\! \text{HRot}(\underbrace{\text{HRot}([\![\mathbf{x}]\!], (i - 1) \cdot r)}_{\text{previous result}}, r)
\end{equation}

Therefore, only three \evks are required for BSGS computation when using the strategy (see Fig.~\ref{fig:BSGS_minimal}), one each for the pre-rotation (Eq.~\ref{eq:prerotation}), the baby-step, and the giant step.

We build on the strategy and propose \emph{\textbf{minimum} key-switching} (Min-KS), which further reduces \evk requirements and has a more broad scope of application.
First, we eliminate the pre-rotation by adjusting and canceling out the rotation amounts between the H-(I)DFT iterations in Alg.~\ref{alg:fft-like-hdft}.
Then, we modify the computational flow of BSGS (see Fig.~\ref{fig:BSGS_multihop}), through which it requires only two \evks per H-(I)DFT iteration.

Second, we generalize Min-KS and apply it to similar computational patterns found in HE workloads.
HE ops such as homomorphic convolution require multiple HRots with rotation amounts exhibiting an arithmetic progression; those are the possible targets of Min-KS.
Also, we aggressively modify the data organization and algorithms to force the rotation amounts to show an arithmetic progression and apply Min-KS to HE ops, such as slot accumulation.

\subsection{On-the-fly limb extension (OF-Limb)}
\label{sec:4_2_of_limb}

We propose \emph{on-the-fly limb extension} (OF-Limb), which generates the limbs of plaintexts used in PMult and PAdd on-the-fly to save off-chip memory access.
The plaintext $\mathtt{P}_{\mathbf{m}^\prime}$ used in $\text{PMult}([\![ \mathbf{m} ]\!] $, $\mathtt{P}_{\mathbf{m}^\prime})$ is precomputed before executing an HE application in the same way as in Eq.~\ref{eq:encode}.
Precomputed $\mathtt{P}_{\mathbf{m}^\prime}$ has the same number of limbs ($\ell + 1$) as the polynomials composing $ [\![ \mathbf{m} ]\!] $ and is loaded from off-chip memory for PMult or PAdd.
However, we identify that it is possible to precompute only the $q_0$-limb of $\mathtt{P}_{\mathbf{m}^\prime}$ and extend it on-the-fly using the following equation so that $\mathtt{P}_{\mathbf{m}^\prime}$ has $(\ell + 1)$ limbs:
 \begin{equation}
     [\mathtt{P}_{\mathbf{m}^\prime}]_\mathcal{C} = \left\{ \text{NTT}([\mathtt{P}_{\mathbf{m}^\prime}]_{q_0} \text{ mod } q_i)\right\}_{q_i \in \mathcal{C}}
 \end{equation}

OF-Limb reduces off-chip memory access of PMult and PAdd to $\sfrac{1}{(\ell + 1)}$ of the original method.
When assuming that the $\mathbf{evk}$s and plaintexts are loaded from off-chip memory and the rest of the working set fits into on-chip memory, plaintexts take up 27.5\% (resp., 40.9\%) of off-chip memory access of H-IDFT (H-DFT).

OF-Limb incurs extra computation.
NTT needs to be performed on the extended limbs of a plaintext to convert them into their evaluation representation.
The increased computation accounts for 22.9\% (resp., 24.1\%) of the entire H-IDFT (H-DFT) computation.
However, we identify that the performance gain from decreased off-chip memory access outweighs the overhead (in Section~\ref{sec:6_3_performance}).

\begin{figure}[t]
\centering
  \subfigure[Homomorphic IDFT\label{fig:idft_intensity}] {\includegraphics[height=1.22in]{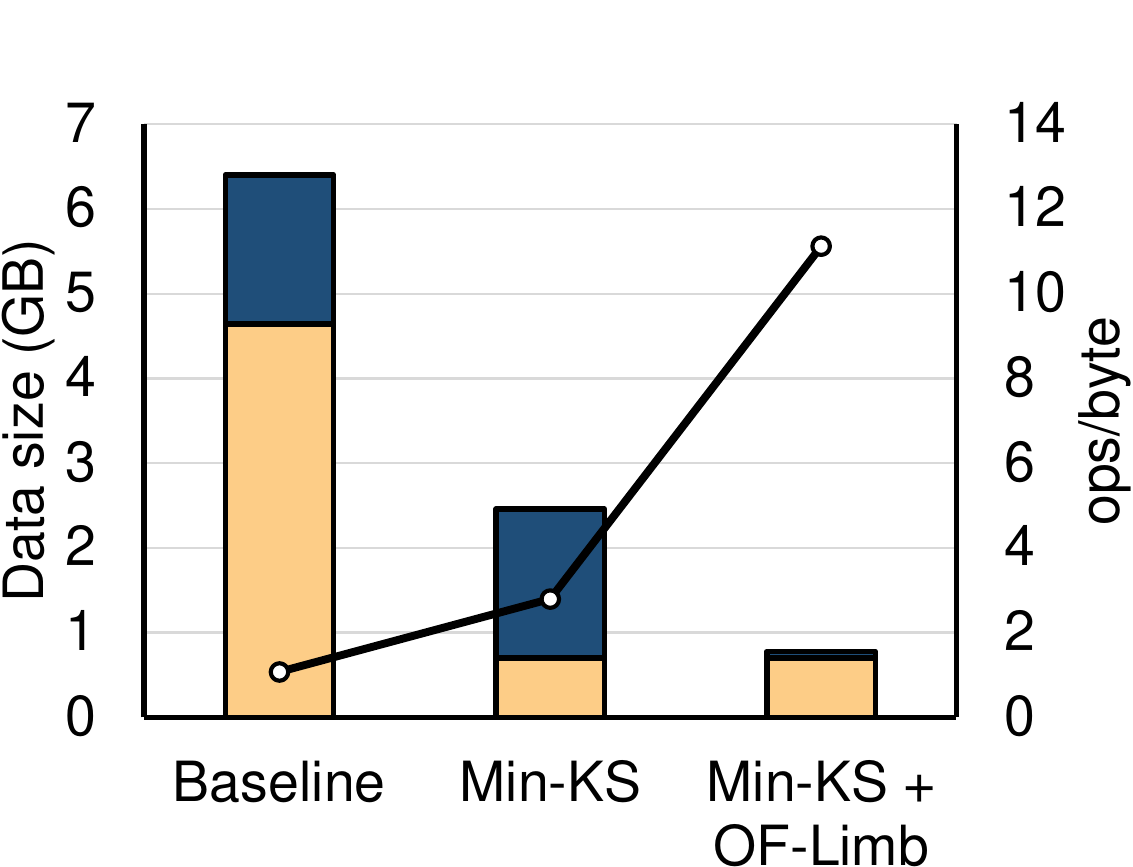}}
  \hfill
   \subfigure[Homomorphic DFT\label{fig:dft_intensity}] {\includegraphics[height=1.22in]{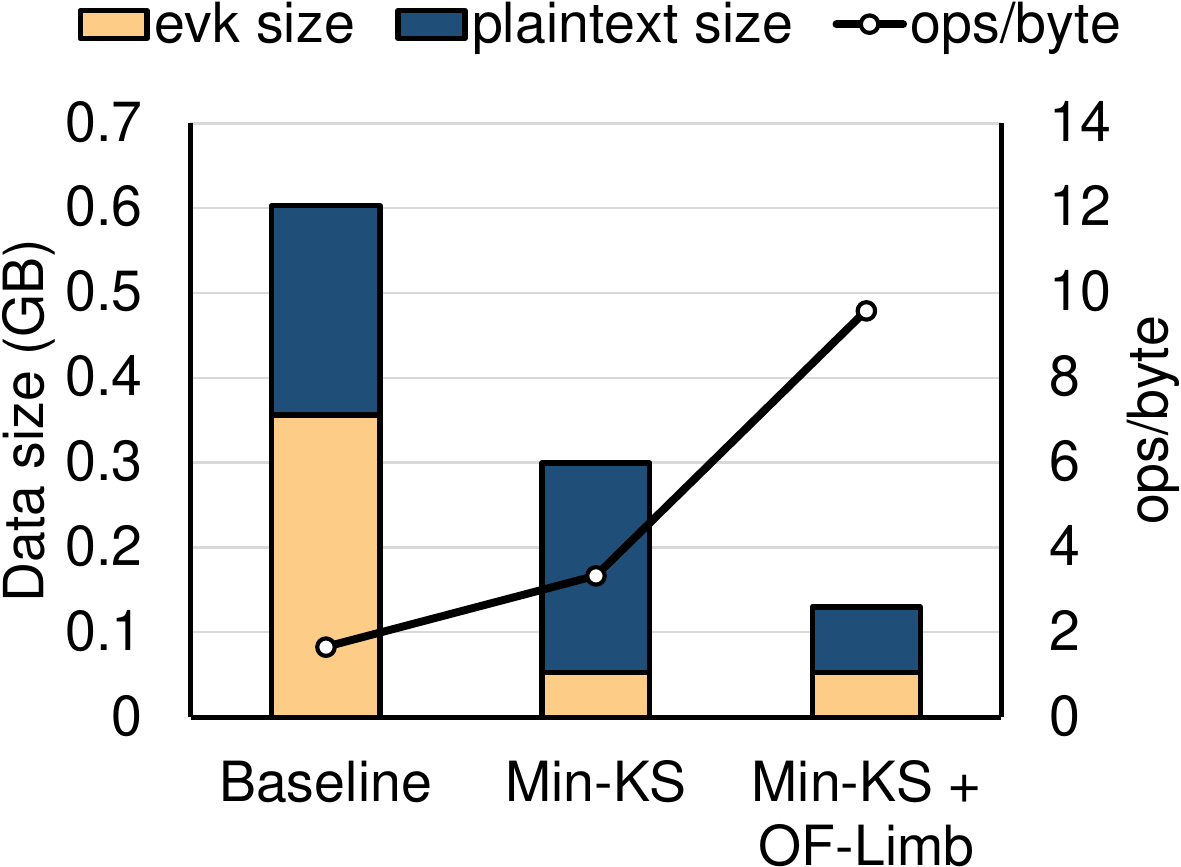}}
    \caption{Data loaded from off-chip memory and arithmetic intensity (ops/byte) change when applying Min-KS and OF-Limb algorithms to (a) homomorphic IDFT and (b) homomorphic DFT.}
    \label{fig:off_chip_access_and_intensity}
\end{figure}

\subsection{Impact of the optimizations on the arithmetic intensity of homomorphic (I)DFT}
\label{sec:4_3_algorithm_impact}

\begin{figure*}[t]
    \centering
    \subfigure[\NAME floorplan\label{fig:floorplan}]{\includegraphics[height=1.41in]{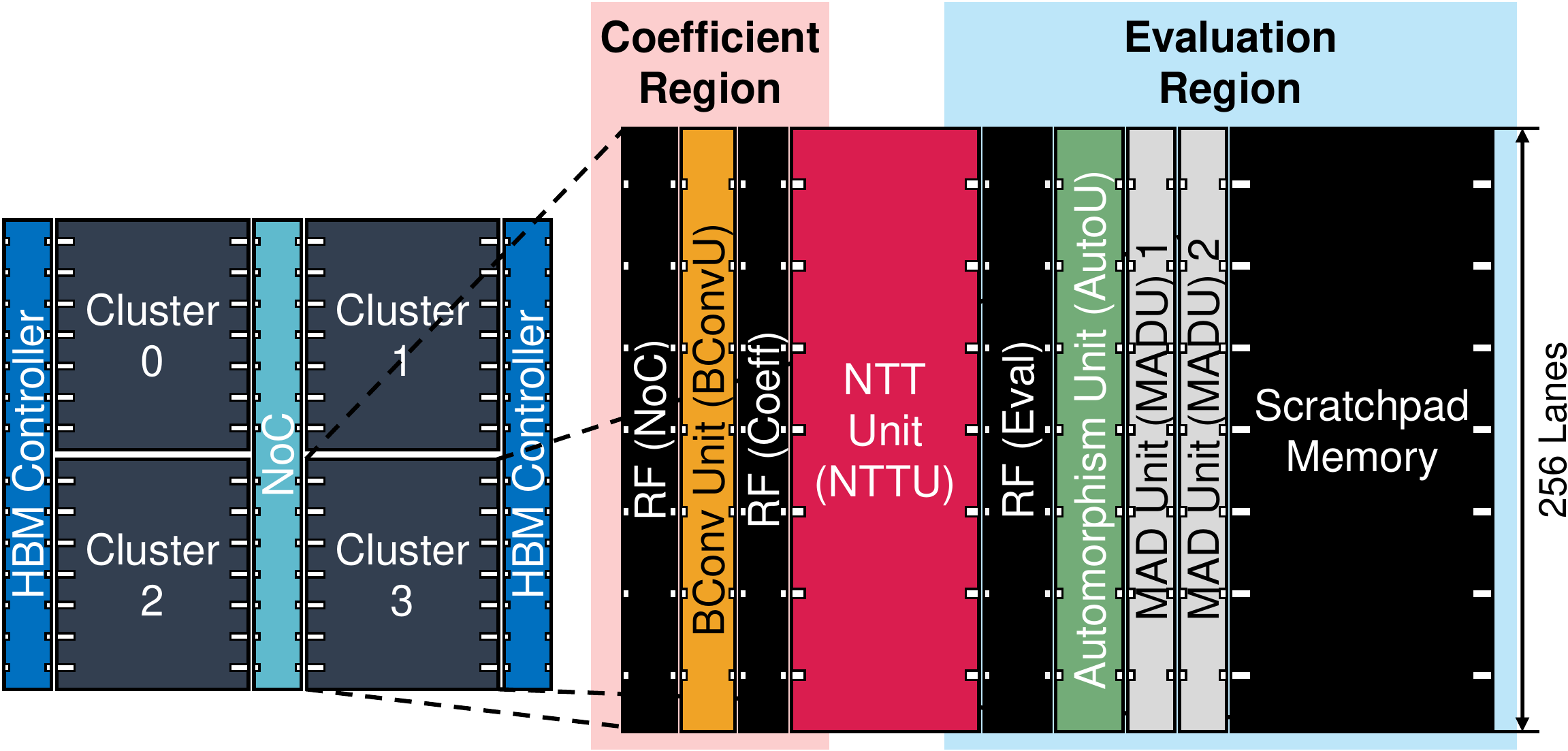}}
    \hfill
    \subfigure[BConv unit\label{fig:bconvu}]{\raisebox{0.032in}{\includegraphics[height=1.18in]{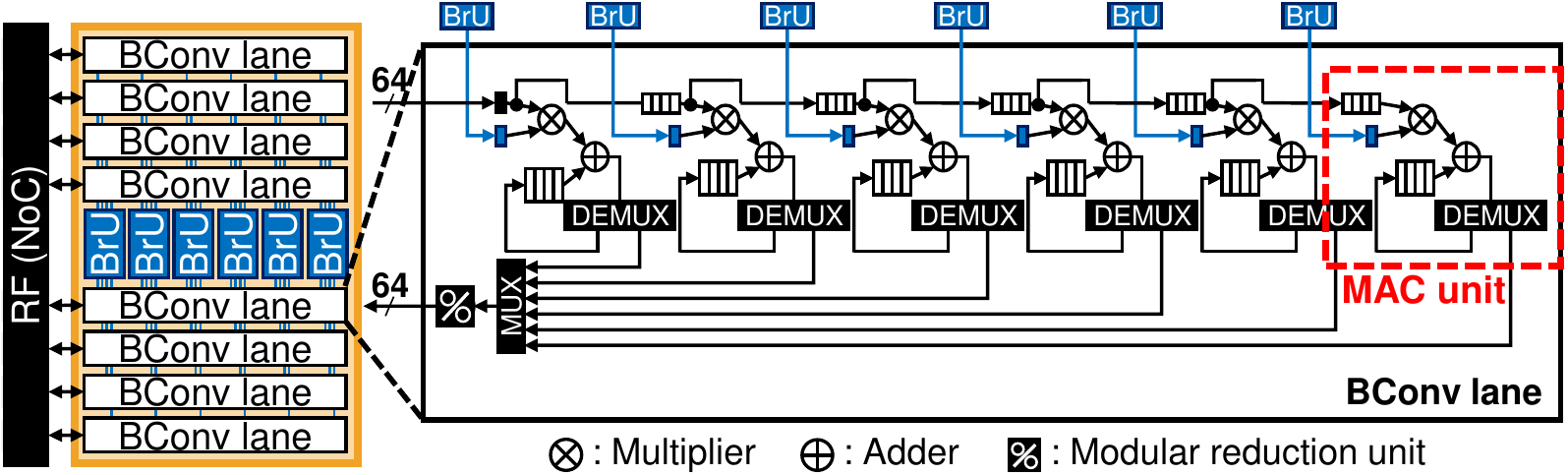}}}
    \caption{(a) Overview of \NAME. \NAME is composed of four clusters. A cluster includes a BConv unit (BConvU), an NTT unit (NTTU), an automorphism unit (AutoU), two multiply-add units (MADUs), scratchpad memory, and distributed register files (RFs). A cluster in \NAME can be divided into two regions: the coefficient region and the evaluation region. (b) Organization of a BConv unit (BConvU). Eight BConv lanes are shown for simplicity. Each lane has six multiply-accumulate (MAC) units. Central broadcast units (BrUs) hold base tables.}
    \label{fig:floorplan_and_bconvu}
\end{figure*}

\begin{figure}[t]
    \centering
    \includegraphics[width=0.99\columnwidth]{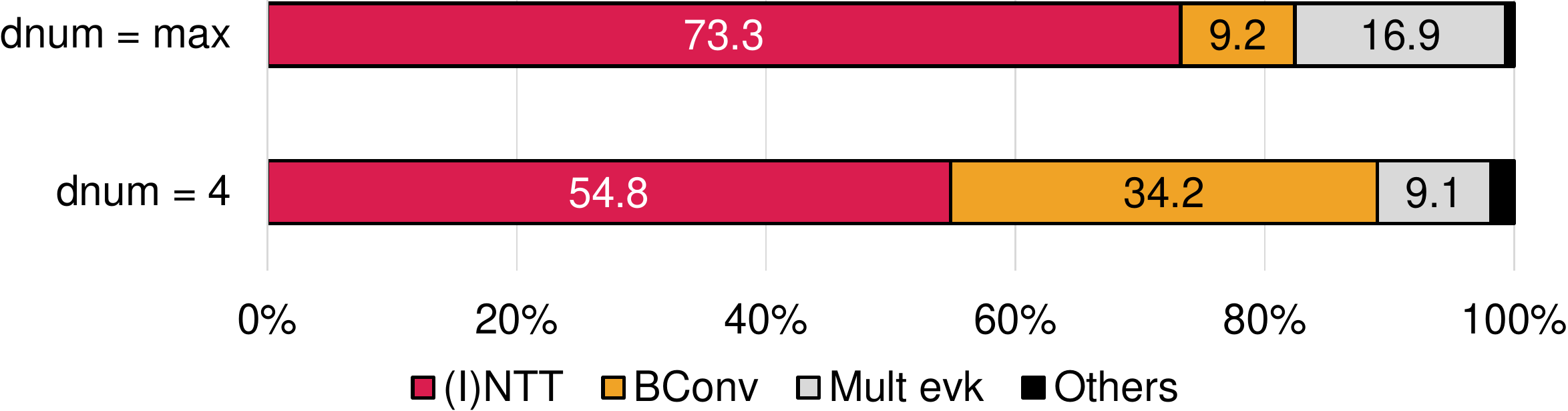}
    \caption{Computational breakdown, in terms of modular mult, of HRot with a ciphertext at the max level for different $\mathtt{dnum}$s. $(N,L) = (2^{16},23)$.}
    \label{fig:ntt_bconv_ratio}
\end{figure}

The two algorithms significantly reduce off-chip memory access, which eventually increases the arithmetic intensity (ops/byte) of H-(I)DFT.
We computed the arithmetic intensity of H-(I)DFT of the baseline algorithm (Section~\ref{sec:3_2_sota_bootstrapping}) with and without applying the two optimizations.
The arithmetic intensity is defined as the number of modular mult ops divided by the off-chip memory access bytes.
Min-KS results in a $2.6\times$ (resp., $2.0\times$) increase in the arithmetic intensity of H-IDFT (H-DFT), and OF-Limb results in an additional $4.0\times$ ($2.9\times$) increase in arithmetic intensity, which reaches 11.1 (9.6) ops/byte.
The algorithms remove 88\% (78\%) of off-chip memory access, as shown in Fig.~\ref{fig:off_chip_access_and_intensity}.

Other optimizations lower the compute cost of H-(I)DFT, such as hoisting~\cite{eurocrypt-2021-efficient}, a technique to merge repeated computation, but they do not reduce the single-use data.
As they are not compatible with Min-KS and only further reduce the arithmetic intensity of the baseline H-(I)DFT algorithm, we excluded them from the analysis.

Min-KS and OF-Limb do not improve performance on conventional systems, such as CPUs and GPUs.
On conventional systems, \evks are always read from the off-chip memory due to the large size of \evks; \evks mostly do not fit in the on-chip storage of conventional systems.
Therefore, reusing \evks, which is enabled by Min-KS, has little impact on the execution time.
Also, conventional systems lack the computational capability to handle the increased arithmetic intensity from OF-Limb cost-effectively.
By contrast, for domain-specific architectures, we can design a system with enough memory capacity and computational capabilities to fully leverage these algorithms.

\section{\NAME architecture}
\label{sec:5_architecture}

We propose \NAME, an architecture-algorithm co-design for FHE that can manage the increased arithmetic intensity and capitalize on the reduced working set by applying the two algorithms.
\NAME utilizes 512MB of on-chip scratchpad memory, enough to hold a couple of $\mathbf{evk}$s and temporary data of HE ops.
Scaling in fabrication technology~\cite{isscc-2017-7nm-sram, isscc-2018-7nm-sram-euv} allowed such massive amounts of on-chip memory; recent domain-specific architectures~\cite{hcs-2021-graphcore, hcs-2021-sambanova, youtube-dojo, isca-2020-groq}, including FHE accelerators~\cite{isca-2022-craterlake, isca-2022-bts}, deployed hundreds of MBs of on-chip memory.
\NAME extensively prefetches \evks, which are often reused by Min-KS.
Also, for the high NTT computation throughput required by OF-Limb, we leverage a high-throughput NTTU similar to that of F1~\cite{micro-2021-f1}.
The NTTU works on a vector of $\sqrt{N}$ elements every cycle, so \NAME also has a form of a vector processor having $ \sqrt{N} \!=\!256$ lanes.
Fig.~\ref{fig:floorplan} shows an overview of \NAME with four clusters, each having an NTTU and other FUs for primary functions of CKKS: a BConv unit (BConvU), an automorphism unit (AutoU), and two multiply-add units (MADUs).

For practical bootstrappable parameters, F1's design targeting the maximum \dnum is no longer applicable.
When using max $\mathtt{dnum}\!=\!24$ under $(N, L) = (2^{16}, 23)$, for instance, the size of an \evk reaches 600MB, which hinders data reuse as it becomes impossible to fit a single \evk into on-chip memory.
Moving to practical parameters using smaller \dnums (see Table~\ref{tab:params}) introduces new data access patterns and computational requirements.
Thus, we design \NAME to handle these new challenges in an FHE accelerator for bootstrappable parameters while using Min-KS and OF-Limb to relieve the off-chip memory bottleneck.

\subsection{Specialized BConv unit for low \dnums}
\label{sec:5_2_bconvu1}

Computational characteristics of HE ops change for lower \dnums.
Prior FPGA and ASIC HE accelerators~\cite{asplos-2020-heax, micro-2021-f1, hpca-2019-roy} targeted small problem sizes with a low $N$, making it inevitable to use the max \dnum to secure more multiplicative levels.
When using the max \dnum, (I)NTT takes up 73.3\% of the total computation (Fig.~\ref{fig:ntt_bconv_ratio}), so accelerating (I)NTT could directly improve the performance.
Therefore, most prior FPGA and ASIC HE accelerators focused on accelerating (I)NTT.
However, for our practical bootstrappable parameters with a large $N$ and a small \dnum, the portion of (I)NTT drops to 54.8\%, and BConv takes up 34.2\% of the total computation.

We design a specialized BConv unit (BConvU) to handle increased BConv computation and relieve the register file (RF) pressure of BConv.
BConv consists of two steps of modular mult.
BConv first multiplies each $p_j$-limb by $(\hat{p}_j^{-1} \text{ mod } p_j)$.
Then, for each $q_i$, BConv multiplies each $p_j$-limb by $(\hat{p}_j \text{ mod } q_i )$, and accumulates the results (see Eq.~\ref{eq:bconv}).
BConvU targets the second step, which accounts for 96\% of the computation of BConv.
The second step is essentially a matrix mult between an $ (\ell + 1) \times \alpha $ matrix composed of $ (\hat{p}_j \text{ mod } q_i)$s, which we refer to as a \emph{base table}, and an $ \alpha \times N$ matrix $[\mathtt{P}_{\mathtt{coeff}}]_\mathcal{B}$.
A na\"ive matrix mult algorithm would require accessing $[\mathtt{P}_{\mathtt{coeff}}]_\mathcal{B}$ along the column direction while $[\mathtt{P}_{\mathtt{coeff}}]_\mathcal{B}$ is stored in memory in row-major order.
To access the matrix along the row direction, our BConvU uses a block matrix mult algorithm~\cite{ics-1992-matmul}, where each of multiple \emph{BConv lanes} processes a column of the polynomial in parallel.

We utilize multiple modular mult-accumulate (MAC) units per lane, forming an output-stationary systolic array~\cite{arxiv-2018-systolic, isca-2021-tpuv4i} to reduce RF pressure.
Fig.~\ref{fig:bconvu} depicts a Bconv lane composed of six MAC units organized as a $1\times6$ systolic array.
A MAC unit passes the input coefficient of a polynomial to the next MAC unit to reuse input data.
A MAC unit is entirely in charge of computing an inner product between a row in a base table and a column in a polynomial.
On each cycle, each of the 256 MAC units on the same horizontal position receives a data element.
These 256 data elements are from the same row of the polynomial, which are multiplied by the same value from the base table.

BConvU features central \emph{broadcast units} (BrUs) that broadcast the base table elements to the MAC units.
In each MAC unit, we time-multiplex mults between a row of the base table and four columns of the polynomial.
This reduces the broadcasting frequency of BrUs to once every four cycles.
We sweep over the number of MAC units to deploy in a lane and choose six, which provides sufficient throughput (see Section~\ref{sec:6_4_alternative}).
By using six MAC units in a BConv lane, the entire BConv process boils down to partitioning the base table into
$6 \times \alpha$-shaped blocks and the polynomial into $\alpha \times (4 \cdot 2^8)$-shaped blocks and performing the block matrix mult between the two.

Each MAC unit has a multiplier, an input buffer to hold four elements, and an output buffer to hold four accumulated results.
Modular reduction logic is placed on the write-back path and is shared by the six MAC units.
Instead, the output buffer of each MAC unit should hold accumulated results in 128 bits.

\subsection{Access-pattern-oriented data distribution}
\label{sec:5_3_coeff_wise_distribution}

BConv has a different data access pattern compared to other functions.
Because an NTTU requires an entire limb of a polynomial to compute using a deep pipeline, and multiple (I)NTTs can be performed on separate limbs of a polynomial, it is natural to distribute a polynomial to multiple clusters \emph{limb-wise}, which is also the data distribution policy F1 used.
However, the limb-wise distribution becomes problematic for performing BConv, where multiple limbs of a polynomial are required to produce an output.
This was not the case when using max \dnum because the input set of limbs was composed of only a single limb when performing BConv, corresponding to $\mathcal{C}_i = \{q_{\alpha i}\} $ or $\mathcal{B} = \{p_0\}$ in Alg.~\ref{alg:key-switching}.

To perform BConv in parallel while distributing data evenly among the clusters, we utilize a \emph{coefficient-wise} data distribution, where $N$ coefficients of a limb in the coefficient representation are distributed evenly across the clusters.
As BConv can be performed in parallel across the coefficients by splitting the $ \alpha \times N $ matrix (polynomial) into $ \alpha \times \frac{N}{4} $ matrices distributed across the four clusters, coefficient-wise distribution is suitable for the data access pattern of BConv.

We devise a data distribution policy that switches between limb-wise and coefficient-wise distribution and corresponding dataflow for performing HE ops in \NAME.
We focus on a BConvRoutine (Alg.~\ref{alg:bconvroutine}).
A polynomial limb passes through an NTTU that executes INTT.
Each cluster prepares an INTT-applied limb and performs an all-to-all data exchange through the network-on-chip (NoC), switching to the coefficient-wise distribution.
When each cluster holds a quarter of $\alpha$ limbs of a polynomial, BConv is executed by a BConvU.
The reverse process is carried out for NTT.
Other than BConv, limb-wise distribution is used.

Switching between limb-wise and coefficient-wise distribution requires all-to-all data communication between clusters.
$ (\alpha + L + 1) \cdot N $ words need to be transferred for every BConvRoutine, requiring a total of $ (\mathtt{dnum} + 2) \cdot (\alpha + L + 1) \cdot N $ words to be transferred in a single key-switching (see Alg.~\ref{alg:key-switching}).
Meanwhile, a limb-wise-only distribution incurs more data communication.
For comparison, we implement the limb-wise-only data distribution by assigning $[\mathtt{P}_\mathbf{m}]_{\mathcal{C}_i}$ to cluster $i$ ($i\! =\! 0, 1, \cdots, \mathtt{dnum}\!-\! 1$) to eliminate data communication for BConvRoutine in Alg.\ref{alg:key-switching}.
This implementation enables the clusters to perform the loop in line \ref{alg:key-switching:line:for} of Alg.~\ref{alg:key-switching} independently.
However, we need to redistribute the data for accumulation (line~\ref{alg:key-switching:line:accum}), which causes more data communication involving $ 2 \cdot \mathtt{dnum} \cdot (\alpha + L + 1) \cdot N $ words when $\mathtt{dnum} > 2$.
We compare the performance of two data distribution policies in Section~\ref{sec:6_4_alternative}.

\begin{figure}[t]
    \centering
    \includegraphics[width=0.88\columnwidth]{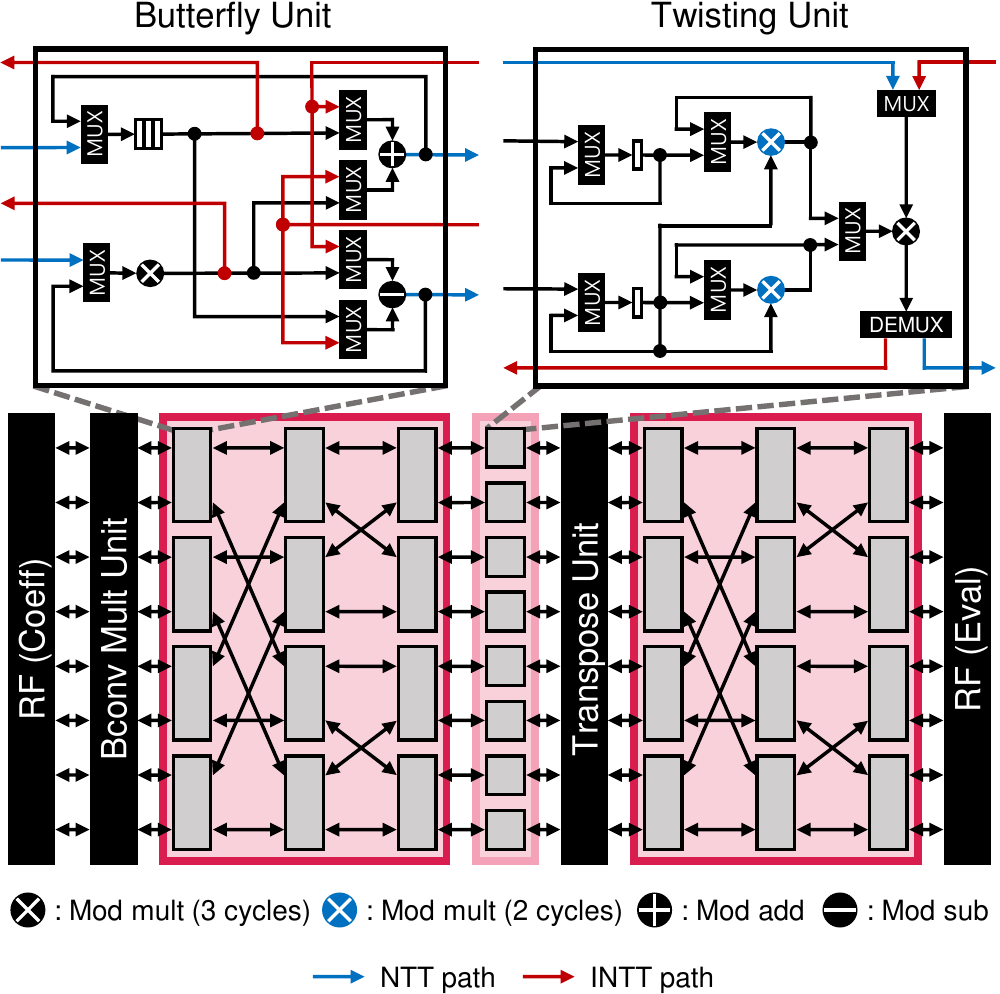}
    \caption{Organization of an NTT unit (NTTU). The logic for 64-point (I)NTT is shown for simplicity.}
    \label{fig:nttu}
    \vspace{-0.15in}
\end{figure}

\subsection{Two regions in a cluster and NTT unit}
\label{sec:6_2_nttu}

Our NTTU, depicted in Fig.~\ref{fig:nttu}, has unique components that focus on minimizing data movement and RF pressure.
First, we distribute RFs across a cluster and organize the NTTU to have the opposite dataflows for INTT and NTT.
In Fig.~\ref{fig:nttu}, data flows rightward for NTT and leftward for INTT.
We partition the RF~\cite{micro-1995-tta, 1996-vector, hpca-2000-register} and place RFs at both ends of the NTTU: RF (Coeff) and RF (Eval).
Then, an NTTU splits a cluster into two regions: the \emph{coefficient region} holding polynomials in the coefficient representation and the \emph{evaluation region} holding polynomials in the evaluation representation.
BConvs and data exchanges on the NoC are executed on the coefficient region, and the other functions are executed on the evaluation region.
Such bidirectional dataflow simplifies the floorplan, which comes down to the organization shown in Fig~\ref{fig:floorplan}, and enables a power-efficient design by reducing data movement that induces expensive wire traversal~\cite{isscc-2014-horowitz, ieee-2001-wire}.
RF (NoC) is added to communicate with the NoC and the BConvU.
Revisiting the dataflow of a BConvRoutine, it can be observed from Fig.~\ref{fig:floorplan} that data movement is minimized by flowing through RF (Eval) $\rightarrow$ NTTU $\rightarrow$ RF (Coeff) $\rightarrow$ NoC $\rightarrow$ RF (NoC) $\rightarrow$ BConvU.
For INTT, NTTU also performs the first step of BConv using the \emph{BConv mult unit} in Fig.~\ref{fig:nttu}.

Moreover, we devise \emph{on-the-fly twisting factor generation} (OF-Twist) on our NTTU, which nearly eliminates memory traffic for loading \emph{twisting factors}.
F1 uses the 4-step FFT in \cite{1990-4stepntt}, which performs $N$-point (I)NTT using 2D-FFT by first performing $\sqrt{N}$-point (I)NTTs, multiplying elements with \emph{twisting factors}, transposing the $\sqrt{N}\times\sqrt{N}$ matrix of elements, and finally performing additional $\sqrt{N}$-point (I)NTTs.
F1 implements the entire process as a single long pipeline by utilizing multiple butterfly units, twisting units, and a transpose unit.
The benefit of this implementation is that the twiddle factors multiplied in the butterfly units stay the same for the whole process.
By contrast, different twisting factors are multiplied for every element, so $N$ twisting factors need to be loaded, of which the amount is as large as the actual data to perform NTT.
We observe that the twisting factors show \emph{geometric progression}.
Thus, we design the twisting units to generate twisting factors on-the-fly by multiplying the common ratios of the geometric progression inside.
Using OF-Twist, we only have to load the starting values and the common ratios, reducing the total data loaded during (I)NTT to roughly half.
Also, OF-Twist can reduce the storage usage for twisting factors by 99\%, saving 30MB of space ($2\cdot (\alpha + L + 1)\cdot N$ words) on our parameter set (Table~\ref{tab:params}).

\begin{figure}[t]
    \centering
    \includegraphics[width=0.8\columnwidth]{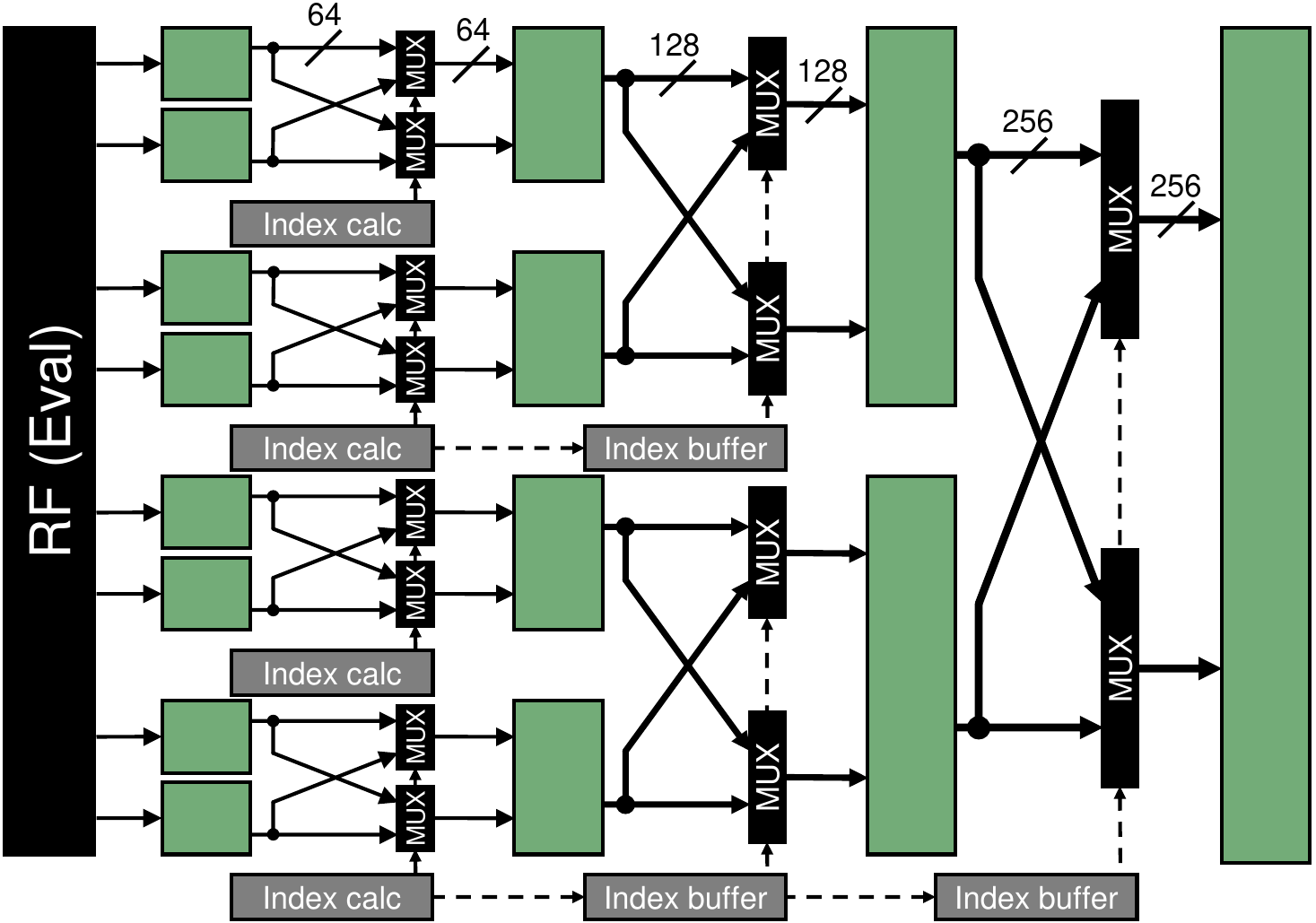}
    \caption{Organization of an automorphism unit (AutoU). The logic for eight words is shown for simplicity. The datapath is shown in solid lines and the control path is shown in dotted lines.}
    \label{fig:autou}
    \vspace{-0.15in}
\end{figure}

\subsection{Automorphism unit (AutoU)}
\label{sec:6_2_autou}

Automorphism moves data according to the mapping $ \psi_r $ in Eq.~\ref{eq:automorphism}. 
Automorphism may look like an irregular permutation between all $N$ coefficients, but it can be performed through internal permutations between $2^8$ coefficients.
Each $2^8$ vector lane consumes a coefficient every cycle, where the $2^8$ consumed coefficients have a stride of $2^8$; i.e., the coefficients with indices $(i, i+1 \cdot 2^8, i+2\cdot 2^8, \cdots , i+(2^8-1)\cdot 2^8)$ are consumed every cycle ($ 0 \leq i < 2^8 $).
After automorphism, the $2^8$ coefficients are again mapped to a set of $2^8$ coefficients with a stride of $2^8$; i.e., $(j, j+1 \cdot 2^8, j+2 \cdot 2^8, \cdots , j+(2^8-1)\cdot 2^8)$ for $ j = \psi_r(i) \text{ mod } 2^8 $.
Thus, automorphism can be done by (i) loading $2^8$ words (one from each lane), (ii) performing an internal permutation of $2^8$ words according to $ \psi_r $, and (iii) storing $2^8$ words (one to each lane).

We devise an automorphism unit (AutoU)\footnote{Our AutoU actually targets a polynomial in the evaluation representation. $\psi_r$ and the data address mapping change for polynomials in the evaluation representation, but the same property still holds.} implementing the internal permutation logic (see Fig.~\ref{fig:autou}).
An AutoU is composed of eight stages, recursively permuting the data according to the $ \psi_r $ index calculation results.
Each stage is pipelined and merges two chunks of data by swapping the order of the two chunks depending on the index calculation results.
The index calculation circuit includes a 16-bit multiplier multiplying the index with precomputed $5^r \text{ mod } N$ (see Eq.~\ref{eq:automorphism}), which can be computed with very little cost.

\section{Implementation}
\label{sec:implementation}

\noindent\textbf{Hardware modeling:} We modeled the FUs of \NAME using ASAP7~\cite{mj-2016-asap7}, a predictive process design kit for a 7nm technology node.
We used FinCACTI~\cite{isvlsi-2014-fincacti} to model the scratchpad and RFs as SRAMs because an SRAM compiler is not included in ASAP7.
We updated the model of FinCACTI to match the information provided by ASAP7, the IRDS roadmap~\cite{whitepaper-2018-irds}, and other published works on 7nm technology nodes~\cite{isca-2021-tpuv4i, iedm-2017-gf7nm, iedm-2016-tsmc7nm, iedm-2017-intel10nm, isscc-2017-7nm-sram, isscc-2018-7nm-sram-euv, isscc-2020-5nm-sram, vlsit-2018-samsung}.
We separately modeled the overhead of expensive semi-global wires (80nm pitch~\cite{intel-2008-wire45nm}), where their length cannot be ignored.
We used \cite{isca-2021-tpuv4i, micro-2017-finegrainedDRAM} to evaluate the HBM memory.
The peak power and area costs of \NAME's components are shown in Table~\ref{tab:area_power}.
Each peak power value is estimated with a usage case where the component can be most active.
\NAME is sized 418.3mm\textsuperscript{2} and consumes up to 281.3W of power.	
	
\noindent\textbf{Performance modeling:} We implemented a cycle-accurate simulator of \NAME to evaluate its performance.
The simulator takes an HE program written at a high level as an input and converts it to a data dependence graph of primary HE functions.
The simulator schedules the functions and data movements while checking the data dependence and structural hazards.
Static scheduling and software-controlled prefetching are possible as HE programs do not feature any dynamic control flows~\cite{micro-2021-f1}.
In that sense, our simulator acts as a VLIW compiler for \NAME.
The simulator collects the utilization rates of the components, combined with the power model, to derive power consumption.

\noindent\textbf{FU implementation and wiring overhead:} All FUs are fully-pipelined and run at 1GHz.
In NTTUs and BConvUs, logic units for modular reduction use Montgomery reduction~\cite{1985-montgomery}, and in MADUs, Barrett reduction~\cite{eurocrypt-1986-barrett}.
Wiring overhead accounts for a significant portion of the FUs in area and power, particularly for NTTUs and AutoUs featuring long wires connecting different lanes in a cluster.
The logic for 4 NTTUs (resp., 4 AutoUs) only accounts for 34.9mm\textsuperscript{2} (0.9mm\textsuperscript{2}) of area and 39.6W (0.3W)  of power.
The rest is attributed to the wiring overhead, including registers for the 1GHz operation of the wires and the repeaters~\cite{ted-2002-repeater}.
The wiring overhead of NTTUs and AutoUs alone accounts for 10\% of the area and 21\% of the peak power of the entire \NAME chip. 

\noindent\textbf{Memory and NoC implementation:} Two 500GB/s HBM2 stacks~\cite{jedec-2021-hbm2, jedec-2022-hbm3} are used, providing a total of 1TB/s off-chip memory bandwidth.
On-chip memory components are single-ported, multiple-banked SRAMs.
The scratchpad memory (128MB per cluster) runs at 1.25GHz and provides a bandwidth of 20TB/s chip-wide.
RF (Coeff) runs double-pumped at 2GHz, while other RFs run at 1GHz.
Distributed RFs (together 19MB per cluster) provide a bandwidth of 72TB/s chip-wide.
NoC is implemented as a simple multiplexer network connecting the same lane from each cluster and provides a modest bandwidth of 8TB/s at 1GHz.

\renewcommand{\arraystretch}{1.0}
\begin{table}[t]
\caption{Area and peak power consumption of \NAME.}
\label{tab:area_power}
\centering
\begin{tabular}{p{2.8cm}rr}
\toprule
\textbf{Component} & \textbf{Area (mm\textsuperscript{2})} & \textbf{Peak power (W)}\\
\midrule
4 BConvUs          & 9.3                                   & 18.9\\
4 NTTUs            & 57.2                                  & 95.2\\
4 AutoUs           & 20.6                                  & 4.6\\
8 MADUs            & 8.9                                   & 24.7\\
Register files     & 42.8                                  & 25.1\\
Scratchpad memory  & 229.2                                 & 54.0\\
NoC                & 20.6                                  & 27.0\\
HBM                & 29.6                                  & 31.8\\
\midrule
\textbf{Sum}       & \textbf{418.3}                        & \textbf{281.3}\\
\bottomrule
\end{tabular}
\end{table}

\section{Evaluation}
\label{sec:6_evaluation}

\subsection{Experimental setup}
\label{sec:6_2_setup}

We first compared the performance of \NAME with other works using a metric called \emph{amortized mult time per slot}~\cite{tches-2021-100x} (\amort).
\cite{tches-2021-100x} proposed \amort as an important metric that allows fair comparison as a proxy compensating for the impact of parameter choice and reported multiplicative throughput in terms of \amort.
\amort adds HMult time for $1, \cdots, L - L_{boot}$ levels ($\textbf{T}_{\text{mult}}(\ell)$) with the bootstrapping time ($\textbf{T}_{\text{boot}}$), then divides it by $L - L_{boot}$ and the number of slots refreshed in bootstrapping ($n$):
\begin{equation}
    \textbf{T}_{A.S.} = \cfrac{\textbf{T}_{\text{boot}} + \sum_{\ell=1}^{L - L_{\text{boot}}} \textbf{T}_{\text{mult}}(\ell)}{L\! -\! L_{\text{boot}}} \cdot \frac{1}{n}
\end{equation}

We also evaluated \NAME using FHE workloads that require bootstrapping.
We did not include tiny LHE workloads as in \cite{micro-2021-f1}, which do not require bootstrapping, as they have little practicality and are not targets of \NAME.
HELR~\cite{aaai-2019-helr} is a simple machine learning workload that trains a binary logistic regression classifier model for the MNIST dataset.
In each iteration, it trains the model with a mini-batch containing 1,024 $14\!\times\!14$-pixel images.
We measured the performance of HELR for 30 iterations and reported the average execution time per iteration.
We also measured the performance of \NAME for more complex workloads.
We performed CNN inference using Lee et al.~\cite{icml-2022-resnet}'s implementation of the ResNet-20 model for the CIFAR-10~\cite{techreport-2009-cifar10} dataset.
The model shows 91.31\% accuracy.
In \cite{icml-2022-resnet}, homomorphic evaluation of convolution layers in the model involves a series of HRots and PMults with weight plaintexts, so we applied both Min-KS and OF-Limb.
Finally, we performed sorting~\cite{tifs-2021-sorting}.
Sorting in HE requires a complex evaluation of non-linear functions approximated by high-degree polynomials.
OF-Limb was applied to every PMult in each workload.

For comparison, we used the state-of-the-art CKKS implementations of \textbf{Lattigo}~\cite{github-lattigo}, \textbf{100x}~\cite{tches-2021-100x} and \textbf{F1}~\cite{micro-2021-f1}, each representing a single-thread CPU, GPU (NVIDIA V100~\cite{techreport-2017-v100}) and ASIC implementation.
We used the reported execution times in \cite{tches-2021-100x, micro-2021-f1}, except for HELR, where \cite{micro-2021-f1} only reported the execution time for a single iteration with 256 images.
We estimated the time for \textbf{F1} to perform 30 iterations of training with 1,024 images by assuming that \textbf{F1} trains 1,024 images over 4 iterations and performs $14\times14\!=\!196$ times of single-slot bootstrapping after each iteration while ignoring other additional costs in favor of \textbf{F1}.
We also estimated execution times for a scaled-up version of \textbf{F1}, \textbf{F1+}, derived by optimistic compensation for the area and fabrication technology gap~\cite{iedm-2017-gf7nm} between \textbf{F1} and \NAME for a fair comparison.
A similar methodology was used in \cite{isca-2022-bts}.
We measured the performance of \textbf{Lattigo} by running the workloads on an Intel CPU system, Xeon Platinum 8358, with 512GB of DDR4-3200 memory.

\subsection{Performance of \NAME}
\label{sec:6_3_performance}

\renewcommand{\arraystretch}{1.1}
\setlength{\tabcolsep}{5.5pt}
\begin{table}[t]
\caption{\amort and HELR execution time of \NAME and prior works}
\label{tab:amort_performance}
\centering
\begin{tabular}{l|rrrr|rr}
\hline
                   & \textbf{Lattigo}  & \textbf{100x} & \textbf{F1} & \textbf{F1+} & \textbf{\NAME} & vs. \textbf{100x}\\
\hline
\hline
\amort ($\mu$s)    & 88                & 8             &  260        & 34           & 0.014          & \textbf{$563\times$}\\
\textbf{HELR} (ms) & 23,293            & 775           & 1,024       & 132          & 7.421          & \textbf{$104\times$}\\
\hline
\end{tabular}
\end{table}
\setlength{\tabcolsep}{6pt}

\begin{table}[t]
\caption{Performance of \NAME for ResNet-20 inference and sorting compared to CPU implementations~\cite{icml-2022-resnet, tifs-2021-sorting}}
\label{tab:workload_performance}
\centering
\begin{tabular}{l|rrr}
\hline
              & CPU                             & \NAME & Speedup\\
\hline
\hline
ResNet-20 (s) & 2,271~\cite{icml-2022-resnet}   & 0.125 & \textbf{18,214$\times$}\\
Sorting (s)   & 23,066~\cite{tifs-2021-sorting} & 1.990 & \textbf{11,590$\times$}\\
\hline
\end{tabular}
\end{table}
\renewcommand{\arraystretch}{1.0}

\begin{figure}[t]
    \centering
    \subfigure[Bootstrapping\label{fig:alg_speedup_a}]{\includegraphics[height=1.72in]{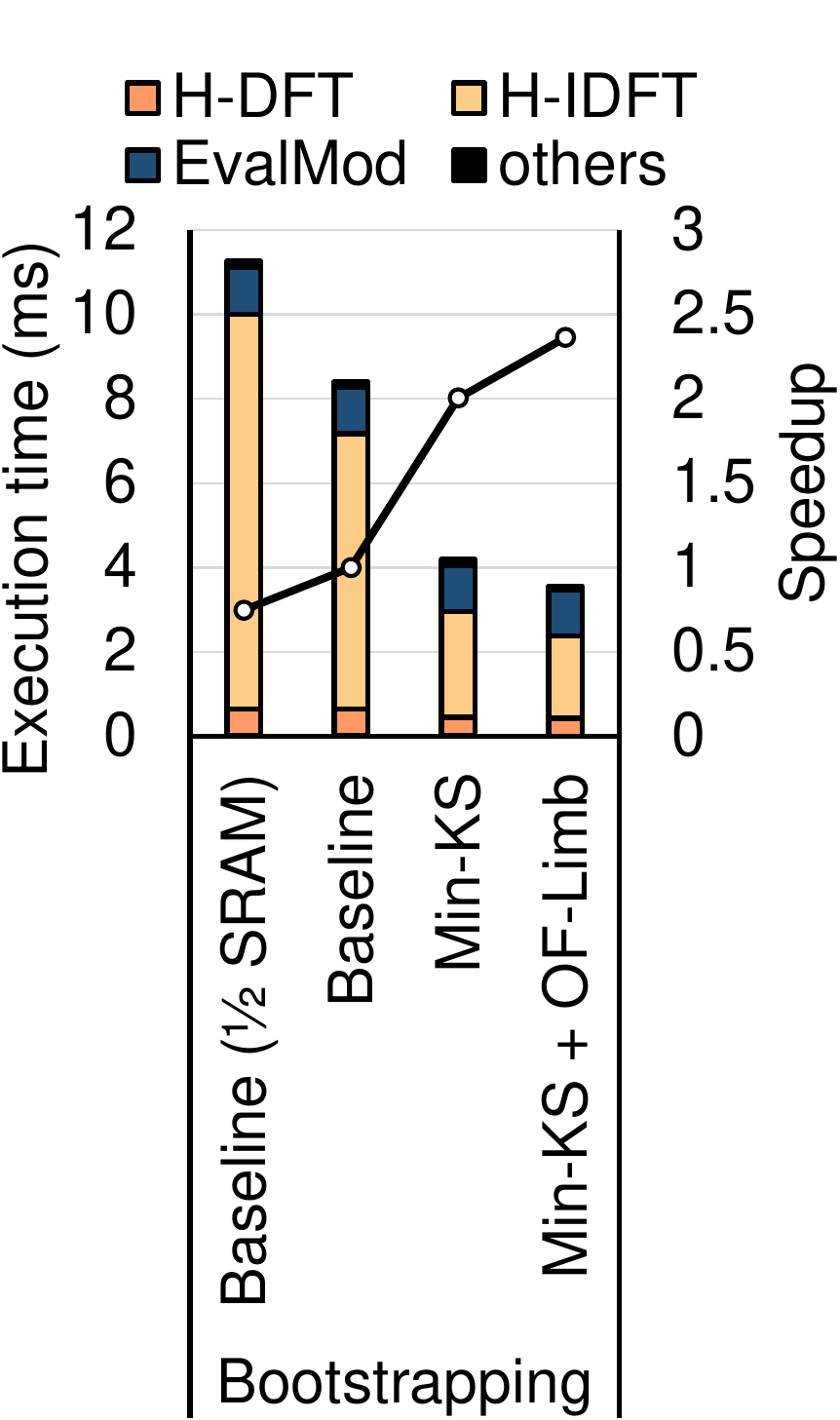}}
    \hspace{0.01in}
    \subfigure[Other workloads\label{fig:alg_speedup_b}]{\includegraphics[height=1.72in]{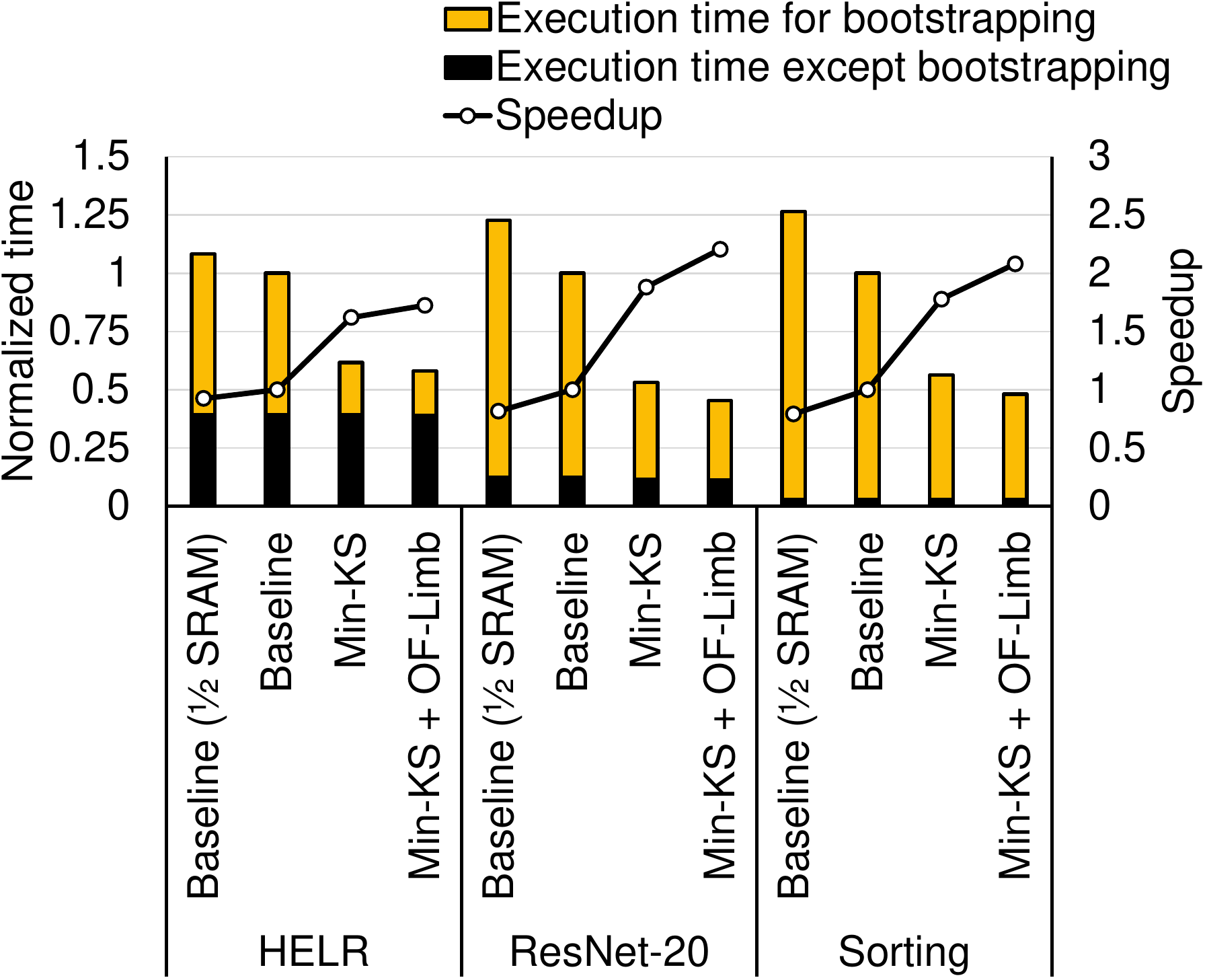}}
    \caption{Execution time and speedup of (a) bootstrapping (single time) and (b) other workloads (normalized) while applying our algorithms incrementally. Execution time and slowdown of \NAME when both algorithms are not applied and also the total scratchpad memory size is halved to 256MB (Baseline (\sfrac{1}{2} SRAM)) are shown as well.}
    \label{fig:alg_speedup}
\end{figure}

\NAME outperforms all other prior works in terms of \amort.
\NAME shows 563$\times$ lower (better) \amort than \textbf{100x}, which performs the best among the prior works (Table~\ref{tab:amort_performance}).
\NAME shows 6,142$\times$ and 2,353$\times$ lower \amort than \textbf{Lattigo} and \textbf{F1+}, respectively.
\textbf{F1} uses single-slot ($n=1$) bootstrapping for CKKS  which is severely limited in throughput, so it shows a similar level of \amort to CPU (\textbf{Lattigo}).

\NAME also outperforms prior works in HELR.
\NAME shows 3,139$\times$, 104$\times$, and 18$\times$ better performance than \textbf{Lattigo}, \textbf{100x}, and \textbf{F1+}, respectively.
The performance gap between \NAME and other works decreases compared to \amort due to reduced slot usage ($n\!=\!256$) in bootstrapping for HELR.
HE operations are comparable to SIMD operations involving all the slots.
Therefore, a fundamental problem in HE is that, for small workloads like HELR that cannot utilize all the slots, throughput is greatly degraded.
Bootstrapping in HELR uses only 256 out of 32,768 slots in \NAME, thus is unable to extract the full capability of \NAME.
This is not the case for \textbf{F1}, which only supports single-slot bootstrapping, in turn showing much enhanced relative performance in HELR compared to other works.
However, for more practical workloads, such as ImageNet~\cite{cvpr-2009-imagenet} with 150,528 pixels in an image, such throughput degradation will be resolved.

\NAME drastically reduces the execution time for complex workloads: ResNet-20~\cite{icml-2022-resnet} and sorting~\cite{tifs-2021-sorting}  (Table~\ref{tab:workload_performance}).
We used the reported performance of CPU implementation in each work as the baseline.
\NAME shows 18,214$\times$ and 11,590$\times$ speedup compared to the baselines.
In particular, \NAME performs a real-time CNN inference in 0.125s.

\noindent\textbf{Impacts of algorithmic optimizations:} Fig.~\ref{fig:alg_speedup_a} shows that our algorithms are effective in reducing the execution time of bootstrapping.
Min-KS improves the performance of H-IDFT (resp., H-DFT) by 2.61$\times$ (1.43$\times$).
OF-Limb further improves the performance by 1.29$\times$ (1.04$\times$), resulting in an aggregate performance improvement of 3.36$\times$ (1.48$\times$).
Overall 2.36$\times$ speedup is achieved in bootstrapping.
Applying Min-KS resulted in a performance gain similar in ratio to the increase in ops/byte (see Fig.~\ref{fig:off_chip_access_and_intensity}), whereas the performance improvement by additionally applying OF-Limb falls short of the increase in ops/byte due to the increased computation.

Our algorithms also result in 1.72$\times$, 2.20$\times$, and 2.08$\times$ speedup for HELR, ResNet-20, and sorting, respectively, as shown in Fig.~\ref{fig:alg_speedup_b}.
Reduced slot usage ($n=256$) in HELR leads to smaller $m$ values in Eq.~\ref{eq:pattern1} and Eq.~\ref{eq:pattern2}, so the overall benefit of Min-KS from reusing \evks is reduced.
Combined with the lower portion of bootstrapping (39.3\%) in HELR, a relatively moderate speedup is achieved in HELR.

The speedup of ResNet-20 attributes partially to the speedup in the convolution layers of ResNet-20.
We additionally applied Min-KS and OF-Limb to the convolution layers, which resulted in a 1.09$\times$ speedup of the execution time, excluding bootstrapping time.
However, for HELR and sorting, only OF-Limb was applied besides bootstrapping, showing a minor performance improvement of less than 1\%.
Other than bootstrapping, these workloads do not feature a computation pattern where Min-KS is applicable.

Simply increasing the on-chip memory capacity has limited performance improvement without algorithmic enhancements.
Fig.~\ref{fig:alg_speedup} shows that, when Min-KS and OF-Limb are not applied, doubling the total scratchpad memory size from 256MB to 512MB results in 1.34$\times$, 1.08$\times$, 1.23$\times$, and 1.26$\times$ speedup for bootstrapping, HELR, ResNet-20, and sorting, respectively.
By contrast, when the two algorithms are applied, the same doubling results in much higher 1.83$\times$, 1.18$\times$, 1.45$\times$, and 1.50$\times$ speedup (parts of the results from Fig.~\ref{fig:sweep_c} and Fig.~\ref{fig:sweep_d}).

\subsection{Alternative designs of \NAME}
\label{sec:6_4_alternative}

\begin{figure}[t]
    \subfigure[Time - boot\label{fig:alternative_design_a}]{\includegraphics[height=1.27in]{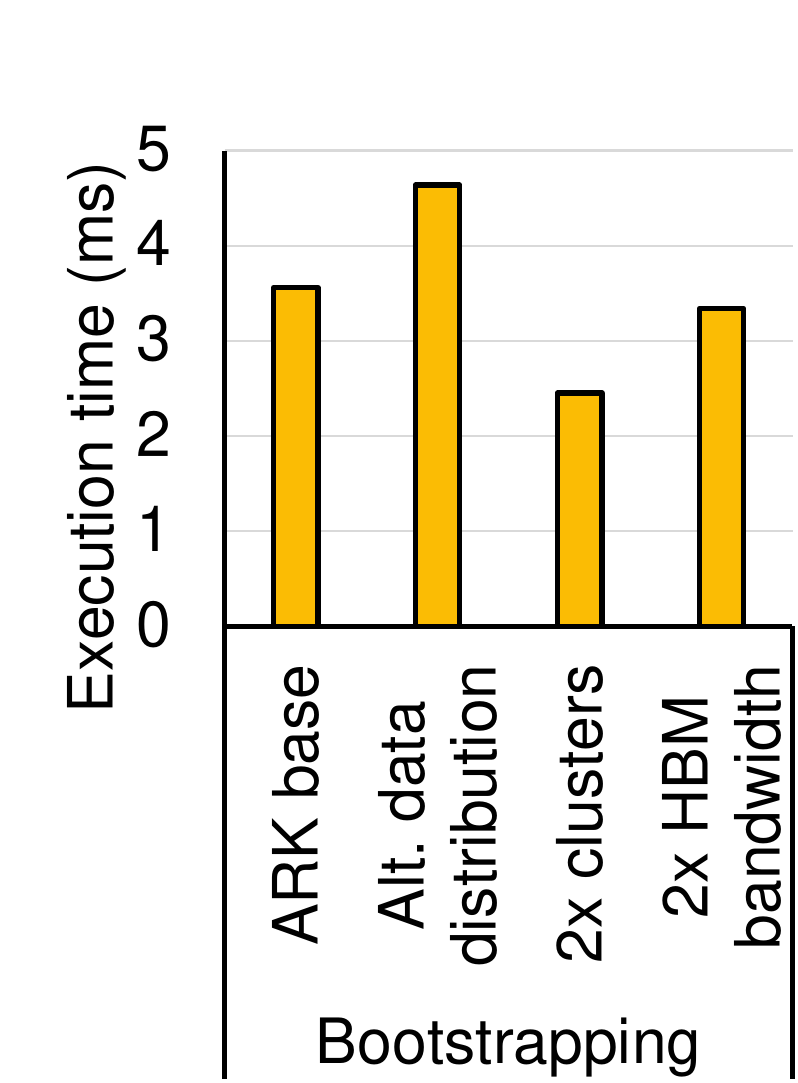}}
    \hspace{0.01in}
    \subfigure[Time - other workloads\label{fig:alternative_design_b}]{\includegraphics[height=1.27in]{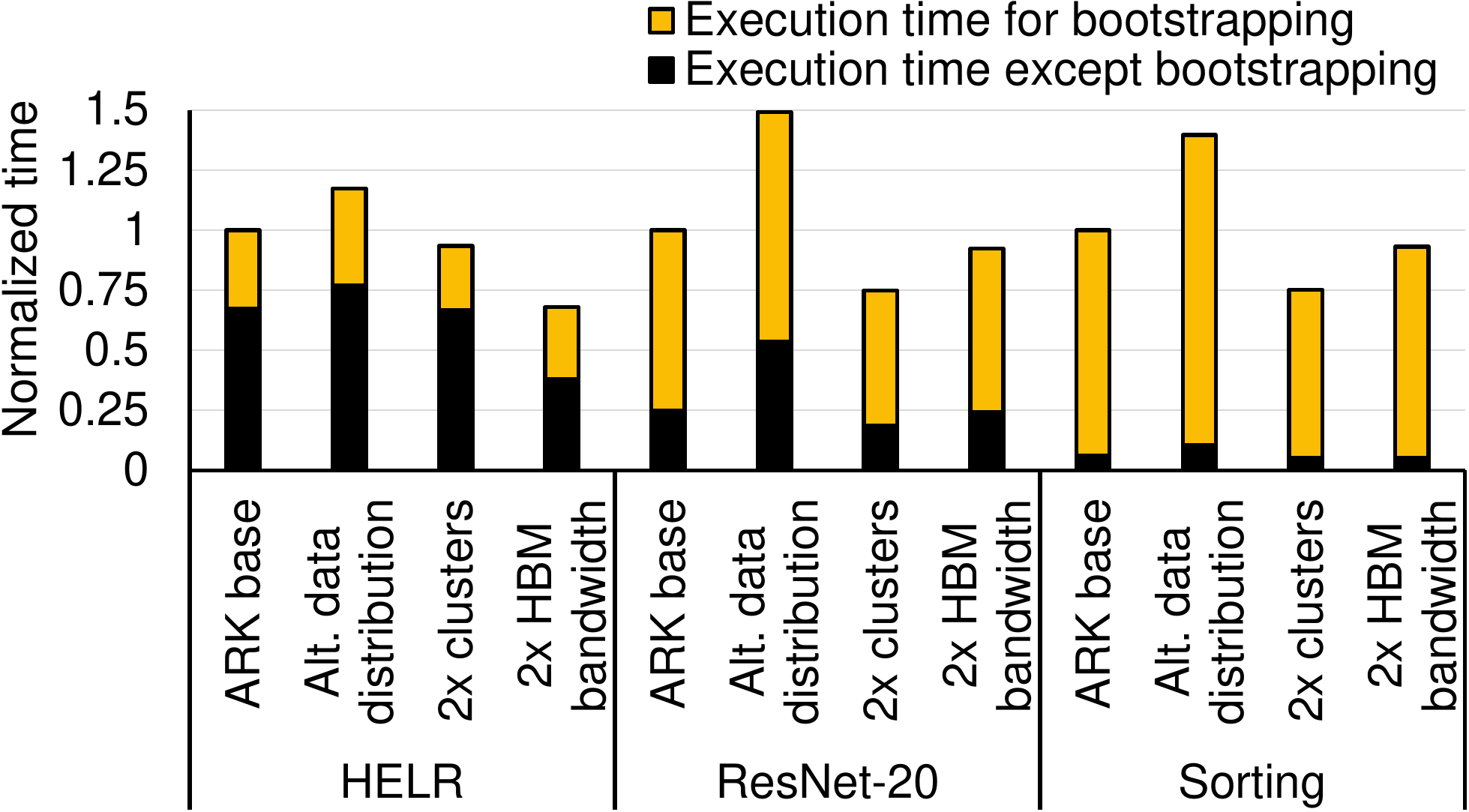}}
    \vskip0.0in
    \subfigure[Power - boot\label{fig:alternative_design_c}]{\includegraphics[height=1.2in]{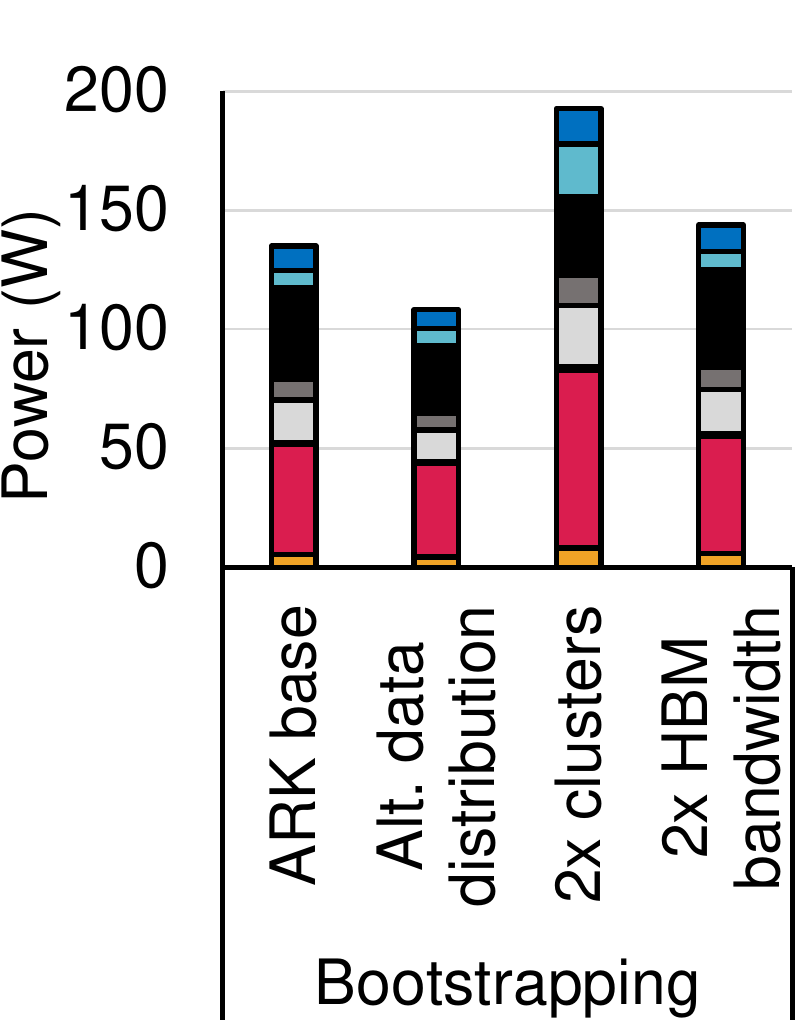}}
    \hspace{0.01in}
    \subfigure[Power - other workloads\label{fig:alternative_design_d}]{\includegraphics[height=1.2in]{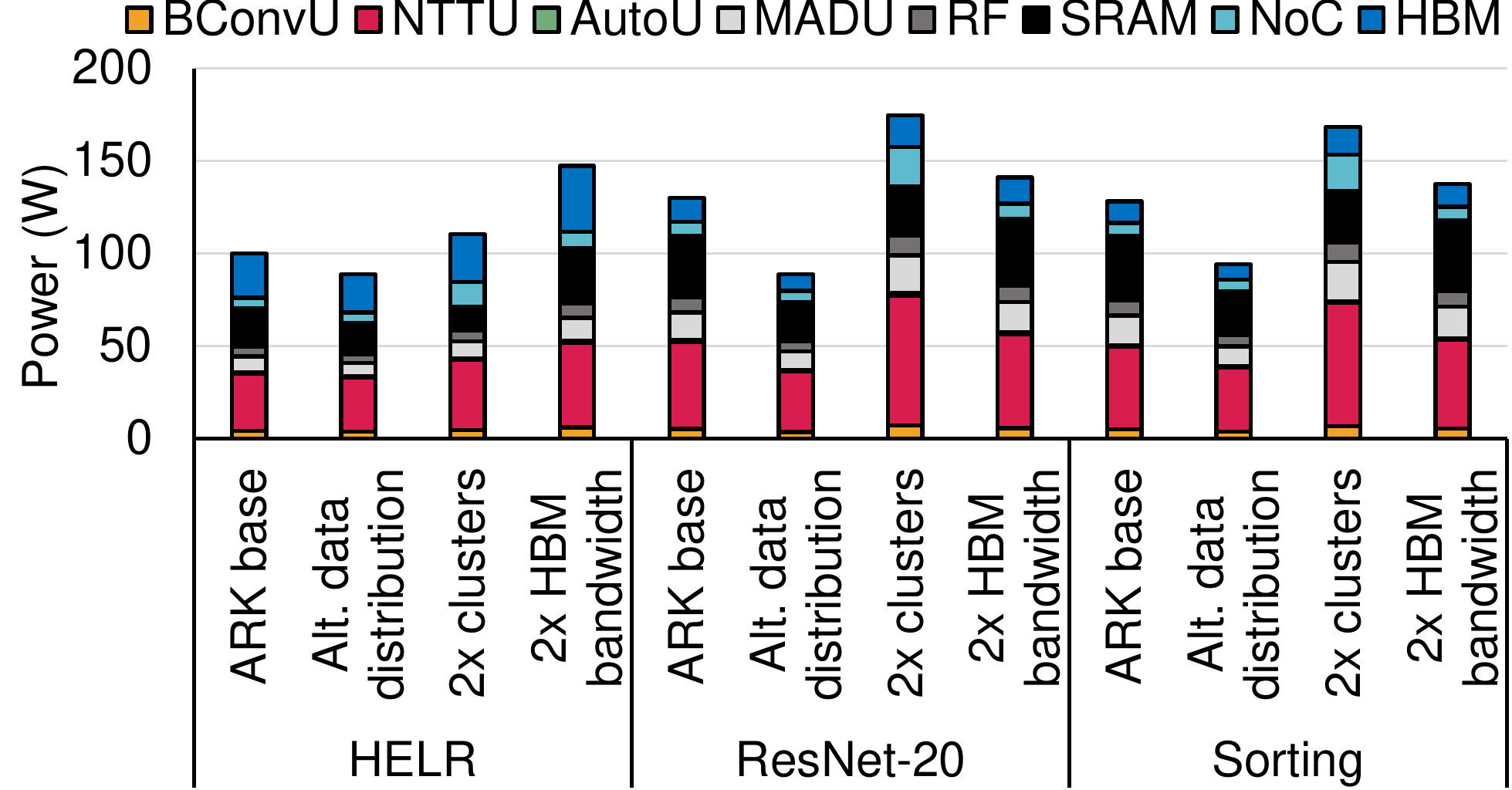}}
    \caption{Execution time and average power for (a)(c) bootstrapping and (b)(d) other workloads with alternative designs of \NAME: \NAME using only limb-wise distribution (Alt. data distribution), eight-clustered \NAME (2$\times$ clusters), and \NAME with 2TB/s main memory bandwidth (2$\times$ HBM bandwidth). Normalized execution time is reported in (b).}
    \label{fig:alternative_design}
\end{figure}

We measured the performance and power of alternative designs of \NAME to assess how each factor of \NAME affects the performance and power.
The results are shown in Fig.~\ref{fig:alternative_design}.
The baseline ARK consumes power ranging from 100W to 135W, 44\% of the peak power in geometric mean (gmean).

\noindent\textbf{Alternative data distribution:} Using limb-wise-only distribution (see Section~\ref{sec:5_3_coeff_wise_distribution}) causes an overall performance drop.
In the limb-wise-only implementation, the data transfer for accumulation after multiplying the evaluation key is the only data transfer.
We designed an NoC suited for the pattern of this data transfer, which deploys adders inside the NoC to support the on-transit accumulation of data, preventing data concentration.
However, even with this additional optimization, performance in bootstrapping, HELR, ResNet-20, and sorting degrades by 0.77$\times$, 0.85$\times$, 0.67$\times$, and 0.72$\times$, respectively.
Power consumption decreases to 0.77$\times$ of the baseline in gmean as the components become less busy.

\noindent\textbf{More clusters:} Because \NAME has eliminated the off-chip memory bandwidth bottleneck, populating more computational logic leads to better performance.
Here we doubled the number of clusters to eight, leading to twice aggregate scratchpad and RF bandwidth while fixing the total scratchpad size to 512MB.
The eight-clustered design can perform bootstrapping 1.45$\times$ faster and thus shows 1.33$\times$ improved performance both for ResNet-20 and sorting.
Due to the lower portion of bootstrapping in execution time, performance improvement in HELR is relatively modest with 1.07$\times$ enhancement. 
However, this performance improvement comes at the cost of 1.29$\times$ more power consumption on average and 1.39$\times$ larger chip area.
In particular, NoC power increases by 2.71$\times$ on average.
When considering the energy-delay-area product (EDAP)~\cite{isca-2008-cacti} to evaluate efficiency, the eight-clustered design shows 1.08$\times$ higher EDAP, suggesting that the baseline \NAME is more efficient.

\noindent\textbf{Doubling HBM bandwidth:} Because the proposed Min-KS and OF-Limb greatly diminish the off-chip memory bandwidth bottleneck, doubling the off-chip memory bandwidth to 2TB/s has little impact on the performance of bootstrapping, which improves by 1.07$\times$.
ResNet-20 and sorting, the performance of which highly depends on bootstrapping, also show a similar degree of performance improvements by $1.08\times$ and $1.07\times$, respectively.
By contrast, HELR has memory-bound parts where multiple HRots with different $\textbf{evk}_{\text{rot}}^{(r)}$s must be performed.
Min-KS is not applicable to those parts as their rotation amounts do not show arithmetic progression.
Therefore, the design with doubled HBM bandwidth shows 1.47$\times$ better performance for HELR.

\begin{figure}[t]
    \subfigure[HELR\label{fig:sweep_a}]{\includegraphics[height=1.22in]{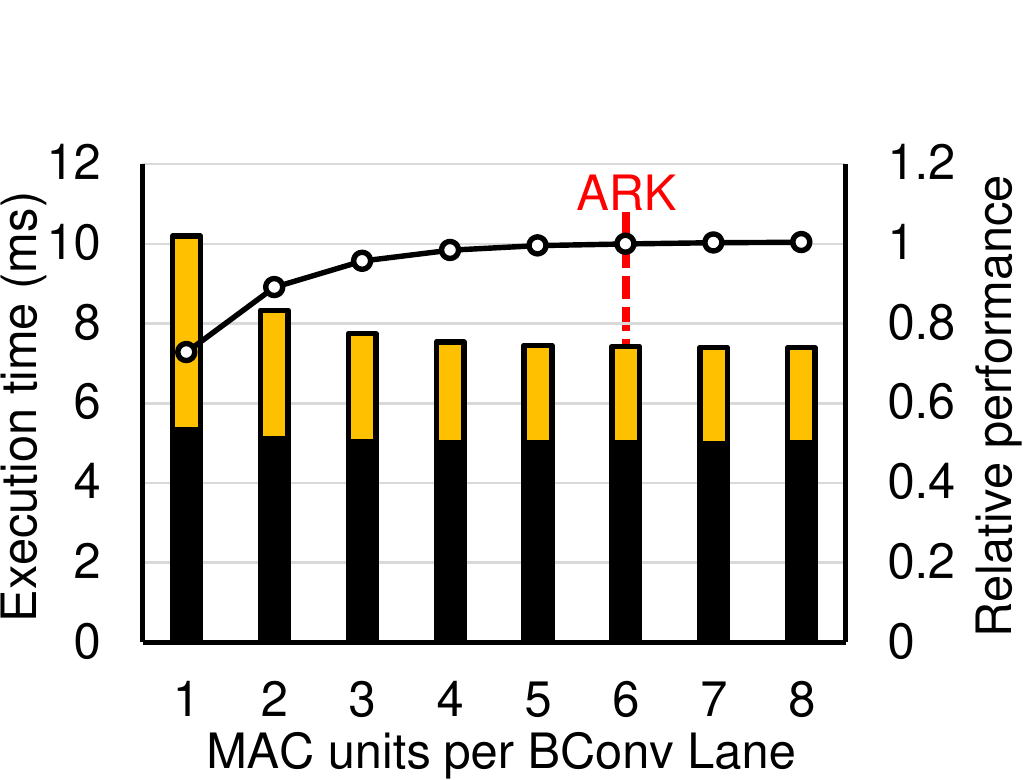}}
    \hspace{0.01in}
    \subfigure[ResNet-20\label{fig:sweep_b}]{\includegraphics[height=1.22in]{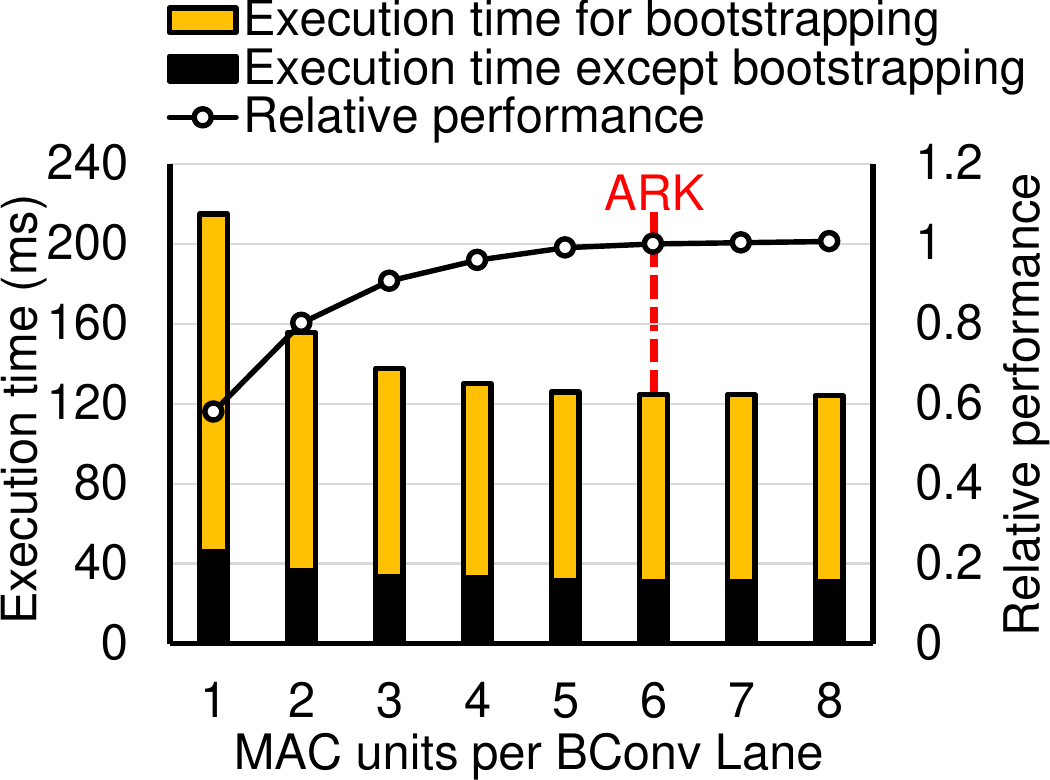}}
    \vskip0.0in
    \subfigure[HELR\label{fig:sweep_c}]{\includegraphics[height=0.992in]{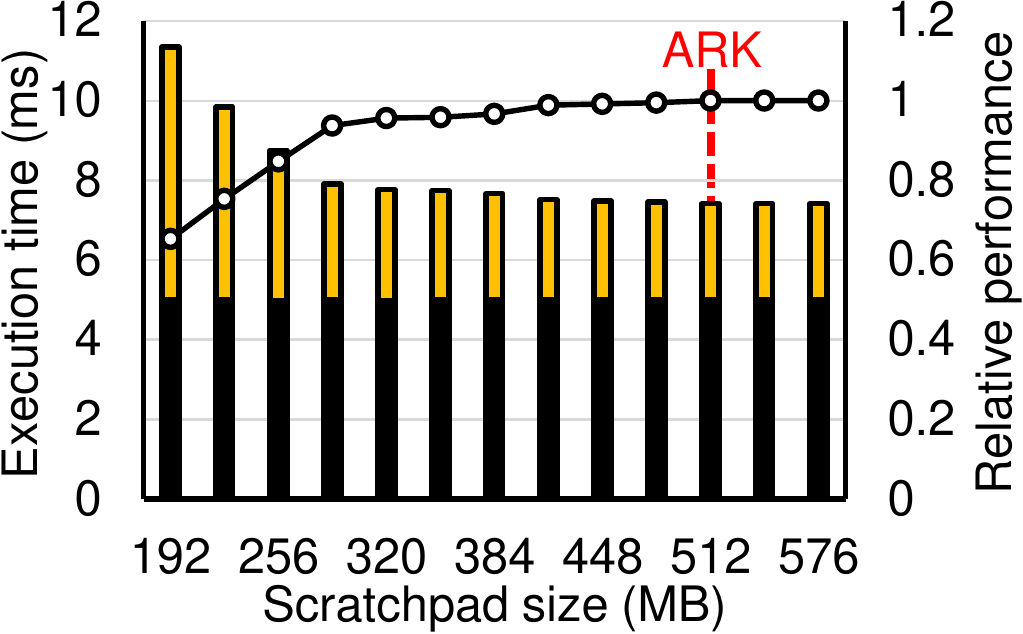}}
    \hspace{0.01in}
    \subfigure[ResNet-20\label{fig:sweep_d}]{\includegraphics[height=0.992in]{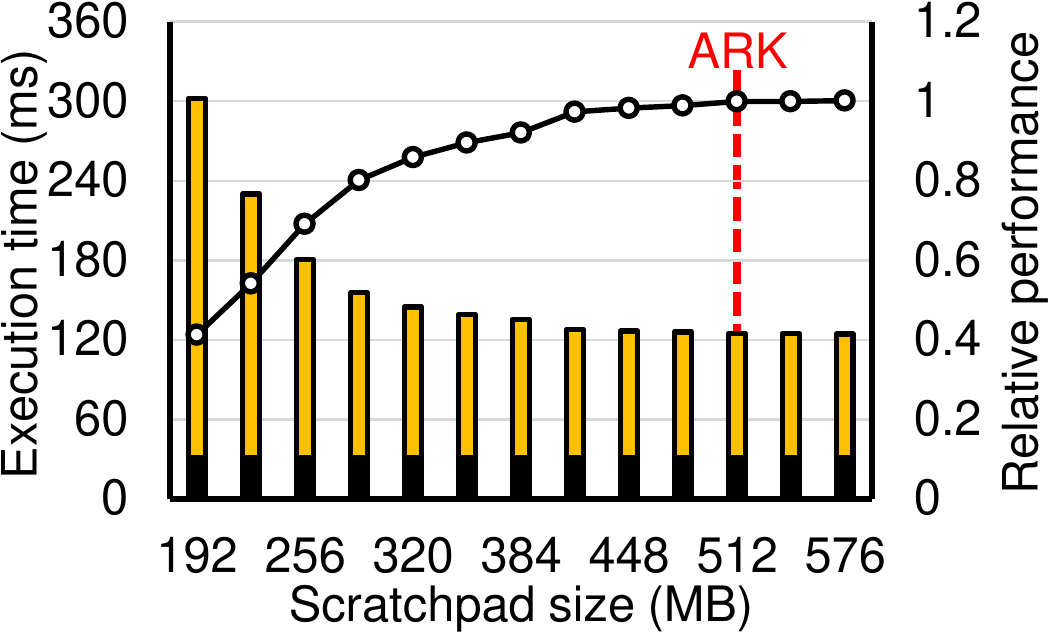}}
    \caption{Performance of \NAME for HELR and ResNet-20 with (a)(b) different numbers of MAC units per BConv lane and (c)(d) different sizes of total scratchpad memory.}
    \label{fig:sweep}
\end{figure}

Fig.~\ref{fig:sweep} shows how performance is affected by the number of MAC units per BConv lane and the size of total on-chip scratchpad memory in HELR and ResNet-20.

\noindent\textbf{Organization of a BConv lane:} Increase in the number of MAC units in a BConv lane from one to six leads to 1.37$\times$ and 1.72$\times$ higher performance in HELR and ResNet-20, respectively, but after six, the performance saturates with less than 1\% enhancement.
HELR is less affected by the number of MAC units as the memory-bound parts of HELR, where multiple data with poor reusability must be loaded, are not accelerated with more computational capability.

\noindent\textbf{Scratchpad memory capacity:} As the amount of total on-chip scratchpad memory increases from 192MB to 512MB, the performance enhances by 1.53$\times$ and 2.42$\times$ in HELR and ResNet-20, respectively, and saturates.
Bootstrapping time is very sensitive to the scratchpad memory size, but other HE ops, being performed at lower multiplicative levels than bootstrapping, need much smaller working sets and are barely accelerated with more scratchpad memory sizes.

\section{Recent FHE accelerators and practicality}
\label{sec:concurrent}

Other FHE accelerators have been recently proposed along with ARK.
Comparisons with those works under 128-bit secure settings are shown in Table~\ref{tab:concurrent}.
CraterLake~\cite{isca-2022-craterlake} is a successor of F1 featuring 256MB of on-chip memory, massive BConv units, and \evk generators which halve the memory traffic for \evks.
CraterLake uses coefficient-wise data distribution.
ARK performs 1.23$\times$ to 2.58$\times$ faster in \amort, HELR, and ResNet-20 compared to CraterLake.
For HELR, \cite{isca-2022-craterlake} only reported performance for training the model with 256 images per iteration, so we estimated the time for CraterLake training 1,024 images over four iterations.
\cite{isca-2022-craterlake} also reported the performance of CraterLake with 350MB of on-chip memory, which shows a speedup of 1.5$\times$ in bootstrapping, but less than 10\% for other workloads.

BTS~\cite{isca-2022-bts} has a tiled architecture composed of 2,048 processing elements (PEs) that communicate through NoCs in the center.
BTS performs an extensive search on optimal HE parameter sets for FHE accelerators under the observation that the loading time of \evks becomes the performance bottleneck.
Then, BTS optimizes the dataflow for HE ops so that they finish within the loading time.
BTS features 512MB of on-chip memory and uses coefficient-wise data distribution. ARK performs 3.19$\times$ to 15.32$\times$ faster in \amort, HELR, ResNet-20, and sorting compared to BTS.

Although huge chip designs are often found in GPUs~\cite{cse-2022-h100, micro-2021-a100, techreport-2017-v100} and domain-specific architectures~\cite{hcs-2021-graphcore, hcs-2021-sambanova, youtube-dojo, isca-2020-groq}, such as Graphcore~\cite{hcs-2021-graphcore} with 896MB of on-chip memory and 823mm\textsuperscript{2} of chip area, FHE accelerators might need to draw more demands to justify such huge chip areas.
CraterLake (472.3mm\textsuperscript{2}), BTS (373.6mm\textsuperscript{2}), and also \NAME (418.3mm\textsuperscript{2}) all employ massive amounts of on-chip memory, consequently utilizing much larger chip areas compared to F1 (151.4mm\textsuperscript{2}).
An increase in fabrication cost, which is superlinear to the chip area~\cite{hennessy-2017-quantitative}, might dilute the benefits of using larger designs despite the also superlinear increase in performance.
Multi-chip modules~\cite{isscc-2020-amdchiplet, jssc-2020-tsmcchiplet} and 3D integration~\cite{ectc-2019-soic, ectc-2020-soic}, as used in recent AMD CPUs~\cite{isscc-2022-zen3} with stacked SRAM memories~\cite{isscc-2022-vcache}, are promising solutions that can lower the fabrication cost by dividing monolithic FHE accelerator designs into chiplet designs.
It is our future work to explore such chiplet FHE accelerator designs.

\renewcommand{\arraystretch}{1.2}
\setlength{\tabcolsep}{6pt}
\begin{table}[t]
\caption{Comparison between \NAME and recent FHE accelerators, CraterLake~\cite{isca-2022-craterlake} and BTS~\cite{isca-2022-bts}}
\label{tab:concurrent}
\centering
\begin{tabular}{l|c|c|c}
\hline
               & \textbf{ARK}                        & CraterLake                 & BTS\\
\hline
\hline
Technology     & \textbf{7nm}                        & 12/14nm                    & 7nm\\
Word size      & \textbf{64-bit}                     & 28-bit                     & 64-bit\\
On-chip memory & \textbf{512MB}                      & 256MB                      & 512MB\\
\hline
\amort         & \textbf{14.3ns}                     & 17.6ns                     & 45.4ns\\
HELR           & \textbf{7.42ms}                     & 15.2ms                     & 28.4ms\\
ResNet-20      & \textbf{0.125s}                     & 0.321s                     & 1.91s\\
Sorting        & \textbf{1.99s}                      & -                          & 15.6s\\
\hline
Area           & \textbf{418.3mm\textsuperscript{2}} & 472.3mm\textsuperscript{2} & 373.6mm\textsuperscript{2}\\
Peak power     & \textbf{281.3W}                     & $>$ 317W                   & 163.2W\\
\hline
\end{tabular}
\end{table}
\renewcommand{\arraystretch}{1.0}
\setlength{\tabcolsep}{6pt}

\section{Related Work}
\label{sec:7_related}

\noindent\textbf{CPU/GPU acceleration of HE:} Prior works have tried to accelerate HE by better use of general-purpose processors: CPUs and GPUs.
\cite{access-2021-demystify} and \cite{wahc-2021-hexl} accelerate HE using CPU SIMD extensions.
Meanwhile, Lattigo~\cite{github-lattigo} exploits the latest algorithmic optimizations~\cite{eurocrypt-2021-efficient, crypto-2018-linear} to accelerate HE ops.
However, the relatively low computational power of CPUs has resulted in limited success.
To overcome the limitations of CPUs, prior works~\cite{tches-2021-100x,access-2020-privft,tetc-2019-bfv,tches-2018-fv} have attempted to accelerate HE by utilizing the high parallel computing power of GPUs.
However, most prior GPU works~\cite{access-2020-privft,tetc-2019-bfv,tches-2018-fv} accelerate part of primitive HE ops or do not include acceleration of bootstrapping.
100x~\cite{tches-2021-100x} accelerates all HE ops including CKKS bootstrapping.
100x mitigates the off-chip memory bottleneck in a GPU by performing kernel fusion, which leads to $242\times$ faster bootstrapping over a CPU.
However, the limited on-chip storage capacity of the GPU hinders kernel fusion for some functions~\cite{iiswc-2020-ntt}.
Yet there are other GPU acceleration works~\cite{github-cufhe, github-nufhe, host-2020-cpuandgpufhe, asap-2017-gpufhew} targeting boolean HE schemes~\cite{jc-2020-tfhe, eurocrypt-2015-fhew}.

\noindent\textbf{FPGA designs for HE:} Prior works have implemented hardware logic for HE ops using FPGAs~\cite{asplos-2020-heax, hpca-2019-roy, fccm-2020-sunwoong-ntt, reconfig-2019-sunwoong-modmult, arxiv-2022-fab}.
However, most of them either do not cover entire HE ops~\cite{fccm-2020-sunwoong-ntt, reconfig-2019-sunwoong-modmult} or target LHE~\cite{asplos-2020-heax,hpca-2019-roy, hpca-2021-cheetah}.
HEAX~\cite{asplos-2020-heax} targets CKKS and accelerates HMult with specialized NTT cores and deep pipelines.
However, it only supports non-bootstrappable, limited parameters.
FAB~\cite{arxiv-2022-fab} is the first FPGA work to accelerate CKKS bootstrapping with a carefully designed datapath for key-switching that minimizes the working set to fully leverage 43MB of its on-chip storage.
FAB shows comparable performance to GPUs~\cite{tches-2021-100x} and ASIC designs~\cite{micro-2021-f1,isca-2022-bts} for some FHE workloads, but its bootstrapping performance is still an order of magnitude lower than recent ASIC FHE accelerators~\cite{isca-2022-bts, isca-2022-craterlake} including \NAME.

\noindent\textbf{ASIC HE accelerators:} A number of ASIC HE accelerators~\cite{hpca-2021-cheetah, micro-2021-f1, isca-2022-craterlake, isca-2022-bts} have been proposed.
Cheetah~\cite{hpca-2021-cheetah} targets privacy-preserving ML using LHE, based on the implementation of Gazelle~\cite{usenixsec-2018-gazelle}, and designs an ASIC accelerator for such workloads.
Gazelle performs ML inference by combining LHE and multi-party computation (MPC)~\cite{sfcs-1986-garbled}.
Such implementation requires frequent server-client communication and needs re-encryption to support complex workloads inducing expensive network overheads.
F1~\cite{micro-2021-f1} is the first ASIC work that reports the performance of CKKS bootstrapping.
By using 64MB of on-chip memory and high throughput FUs, including NTTU, F1 improves the performance of primitive HE ops; however, it does not target the parameters suitable for practical bootstrapping.
Recently proposed BTS~\cite{isca-2022-bts} and CraterLake~\cite{isca-2022-craterlake} are state-of-the-art ASIC FHE accelerators that fully consider appropriate parameters for CKKS bootstrapping.

\section{conclusion}
\label{sec:8_conclusion}

In this paper, we present a set of fabrication-technology-aware algorithmic optimizations and hardware design for fully homomorphic encryption (FHE).
To mitigate the off-chip memory bandwidth bottleneck of bootstrapping, we present the following key algorithmic enhancements: \emph{minimum key-switching} and \emph{on-the-fly limb extension}.
These two optimizations significantly reduce off-chip memory access, fully reaping the high computational capabilities of domain-specific architectures.
We propose \NAME, an accelerator addressing new computation and data movement constraints of FHE through an efficient functional unit (FU) microarchitecture that minimizes on-chip data movement.
\NAME features a data-access-pattern-aware data distribution policy and a memory organization simplifying dataflow, enabling streamlined execution across the FUs.
Overall, \NAME provides two to four orders of magnitude higher performance than prior works.
This enables \NAME to provide real-time encrypted CNN inference, showing 0.125 seconds of inference time for the ResNet-20 model.

\section*{Acknowledgment}
The authors thank Jaewan Choi, the anonymous reviewers, and the shepherd for their constructive comments.
This work was supported by the National Research Foundation of Korea (NRF) grant (No. 2020R1A2C2010601) and Institute of Information \& communications Technology Planning \& Evaluation (IITP) grants (No. 2020-0-00840, and No. 2021-0-01343) funded by the Korean government (MSIT).
The EDA tool was supported by the IC Design Education Center (IDEC), Korea.
Jongmin Kim is with the Interdisciplinary Program in Artificial Intelligence, Seoul National University (SNU).
Sangpyo Kim is with the Department of Intelligence and Information, SNU.
Jung Ho Ahn, the corresponding author, is with the Department of Intelligence and Information,
the Interdisciplinary Program in Artificial Intelligence,
and the Research Institute for Convergence Science, SNU.

\bibliographystyle{IEEEtranS}
\balance
\bibliography{refs}

\end{document}